\documentclass[10pt]{elsarticle}
\usepackage{amsfonts}
\usepackage{amsmath}
\usepackage{amsthm}
\usepackage{amssymb}
\usepackage{mathrsfs}
\usepackage{graphicx}
\usepackage[bf,center]{titlesec}
\usepackage{verbatim}

\newtheoremstyle{break}
  {}
  {}
  {\itshape}
  {}
  {\bfseries}
  {}
  {\newline}
  {}

\theoremstyle{break}

\newtheorem{cor}{Corollary}[section]

\newtheorem{lem}{Lemma}[section]

\newtheorem{tm}{Theorem}[section]
\newtheorem{asp}{Assumption}[section]
\renewcommand{\iint}{\int \!\!\!\! \int}

\long\def\symbolfootnote[#1]#2{\begingroup%
\def\thefootnote{\fnsymbol{footnote}}\footnote[#1]{#2}\endgroup}
\newcommand{\arctanh}{\mathrm{artanh}}

\title{On Linear Landau Damping for Relativistic Plasmas via Gevrey Regularity}
\author{Brent Young \\ Universit\"at zu K\"oln \\ bojy77@gmail.com}

\begin{document}
\maketitle

\begin{abstract}
We examine the phenomenon of Landau Damping in relativistic plasmas via a study of the relativistic Vlasov-Poisson system (both on the torus and on $\mathbb{R}^3$) linearized around a sufficiently nice, spatially uniform kinetic equilibrium.  We find that exponential decay of spatial Fourier modes is impossible under modest symmetry assumptions.  However, by assuming the equilibrium and initial data are sufficiently regular functions of velocity for a given wavevector (in particular that they exhibit a kind of Gevrey regularity), we show that it is possible for the mode associated to this wavevector to decay like $\exp(-|t|^{\delta})$ (with $0 < \delta < 1$) if the magnitude of the wavevector exceeds a certain critical size which depends on the character of the interaction.  We also give a heuristic argument why one should not expect such rapid decay for modes with wavevectors below this threshold.
\end{abstract}

\tableofcontents

\section{Introduction}

One of the more striking results for a non-relativistic one-component Coulomb plasma with a (uniform) neutralizing background charge density $\overline{\rho}$ in unbounded space, as modeled by the Vlasov-Poisson system
\begin{equation}
\textrm{VP}^+: \left\{ \begin{array}{l} \partial_t f + p\cdot \nabla_q f + E\cdot\nabla_p f = 0\\
\nabla\cdot E = 4 \pi (\rho - \overline{\rho})\\
\rho(t,q) = \int f(t,q,p)\;d^3p
 \end{array}\right. ,
\end{equation}
is the phenomenon of \emph{Landau Damping}.  As first noted by Landau in 1946 \cite{L46}, solutions to the linearization of VP$^+$ about a uniform, Maxwellian equilibrium can exhibit exponential decay of Fourier modes associated to non-zero wavevectors in their spatial distributions, $\rho$ (the electric field will also decay exponentially fast in such situations).  As such, the linearized system seems to exhibit a \emph{time-irreversible} behavior (exponential decay to a constant background).  What makes this result so surprising at first glance is that VP$^+$ itself is \emph{time-reversible} (and so also is its linearization).  By contrast, traditional approach-to-equilibrium results (e.g., plasmas described by the Boltzmann equation tend to the uniform Maxwellian background) all involve irreversible equations describing dissipative systems where some Lyapunov functional is decreasing as $t$ increases (such as the negative entropy functional for the Boltzmann equation).  Due to the form of the Vlasov equation however, any reasonable Lyapunov functional will be preserved under the evolution!  Many different physical mechanisms have been put forward to explain this apparent paradox, but from a purely mathematical perspective,  the decrease in amplitude of $\rho$ is paid for at the expense of increasing derivative norms for $f$.  The increasing filamentation in phase space (accompanied by increasingly higher frequency oscillations) is such that it averages out in the marginal distribution $\rho$.  Hence, from a mathematical point-of-view, Landau Damping is a kind of weak convergence result.  For a nice introduction to these ideas, see \cite{V10}.

Once we move from a mode-by-mode analysis to a full treatment of the linearized system,  things become much trickier.  Since the zero Fourier mode is always preserved in time (as this represents the total charge), it is reasonable to expect that only modes associated to wavevectors well separated from zero can exhibit \emph{uniform} exponential decay in time.  On the 3-dimensional torus of size $L$, this is a given since the smallest non-zero wavevector has magnitude $L^{-1}$.  For plasmas distributed on the entirety of $\mathbb{R}^3$, we can only hope that modes above a certain threshold will decay uniformly exponentially fast.  Indeed,  Glassey and Schaeffer \cite{GS94} have shown that for VP$^+$ in one spatial dimension linearized about a uniform kinetic equilibrium, $f_0(p)$, the best overall decay rate (as measured by $L^2$-norm) one can hope for is $\mathcal{O}(t^{-1})$.  For the Maxwellian equilibrium, the decay rate is only like $(\ln t)^{-3/2}$ (in three spatial dimensions, they also show the decay for the Maxwellian is like $(\ln t)^{-5/2}$), and for $f_0$ radially decreasing and compactly supported, there can be no decay at all.  In general, the faster $f_0'$ limits to zero as $|p| \to \infty$, the slower $\rho$ tends to zero in $L^2$-norm.  Hence, confinement of the plasma (as in the torus) seems essential for true exponential decay.

For many years, this damping phenomenon was known rigorously only for the linearized system.  Recently, Mouhot and Villani \cite{MV11} have succeeded in showing that sufficiently regular solutions to both the fully non-linear VP$^{+}$ and VP$^{-}$ systems on the torus do indeed exhibit the damping phenomenon exponentially in time; here, VP$^-$ is the gravitational analog of the Coulombic VP$^+$.  More precisely, they show that for an analytic kinetic equilibrium, $f_0 = f_0(p)$, satisfying certain stability criteria (along with constraints on the sizes of its derivatives and Fourier transform) there is an $\epsilon >0$ so that all initial data, $f_i = f_i(p,q)$, within $\epsilon$ of $f_0$ in an appropriate norm limit exponentially fast in $t$ to a spatially uniform state as $t \to \infty$.

When we inquire about Landau Damping for relativistic plasmas, we find much less information in the literature.  In 1994, Schlickeiser \cite{S94} examined the phenomenon for mono-charged, relativistic plasmas close to the spatially uniform J\"uttner distribution (which is the relativistic version of the Maxwellian profile).  Working with an expression for the plasma conductivity tensor linearized about this relativistic equilibrium (derived earlier by Trubnikov),  he found that there is a temperature-dependent critical magnitude, $k_c$, so that ``superluminal oscillations undergo no Landau damping,'' (see the abstract of \cite{S94}) corresponding to wavevectors of magnitude below the critical value (it is our reading of the paper \cite{S94} that ``Landau damping'' is meant in the strict sense of exponential decay).  This is in sharp contradistinction to the situation for non-relativistic plasmas linearized about the Maxwellian where no such critical $k_c$ is to be found.  As such, mono-charged relativistic plasmas may not exhibit exponential decay to the uniform equilibrium on the torus even for very nice initial data (depending, of course, on the ambient temperature and size of the torus).

It is the purpose of this paper to begin a rigorous examination of the behavior of relativistic plasmas in the spirit of Mouhot and Villani.  As a first step, we study the behavior of Fourier modes for the \emph{relativistic Vlasov-Poisson system} (rVP which we define below) linearized about a sufficiently nice, spatially uniform kinetic equilibrium.  Since at this level we work only on a mode-by-mode basis, many of our computations hold equally well in the full space as on the torus.  As such, we provide results in both cases.

Incidentally, ``relativistic Vlasov-Poisson'' may sound like a strange mix of relativistic and non-relativistic physics, but in fact it is a special case of the \emph{relativistic Vlasov-Maxwell system} (rVM):
\begin{equation}
\textrm{rVM:} \left\{ \begin{array}{l} \partial_t f + v(p)\cdot \nabla_q f + \sigma\left(E+v(p)\times B\right)\cdot\nabla_p f = 0\\
\partial_t E = \nabla \times B - j\\
\partial_t B = -\nabla\times E,\\
\nabla\cdot E = 4\pi (\rho - \overline{\rho}), \;\; \nabla \cdot B = 0\\
\rho(t,q) = \int f(t,q,p)\;d^3p\\
j(t,q) = \int v(p)f(t,q,p)\;d^3p
 \end{array}\right.,
\end{equation}
which describes the evolution of a mono-charged, dilute (i.e. collisionless) plasma with phase-space distribution function $f >0$ (the magnitude of charge for the particles comprising the plasma is given by $\sigma >0$); these equations are understood to be in the rest frame of the neutralizing background (otherwise, one would need to add a corresponding background current $\overline{j}$).  The relativistic velocity in terms of the momentum is given by
\begin{equation}
v(p) = \frac{p}{\sqrt{1+|p|^2}},
\end{equation}
in units where the speed of light and the mass of the particles in question are both equal to 1.  For an excellent introduction to this system see \cite{R04}.  \symbolfootnote[2]{In particular, this paper gives a nice review of the existence of global weak solutions to rVM.  As for results on decay rates, there are a few results (at least in certain special cases).  In 2010, Glassey, Pankavich, and Schaeffer \cite{GPS10} showed that there are solutions to rVM in 1.5 dimensions (i.e. one spatial dimension and two momenta dimensions) for which the spatial distribution of charge exhibits no decay in $t$.  In fact, all $L^p$-norms of the distribution for $p\in [1,\infty]$ are bounded below by a constant which is independent of $t$.  They also show that there are no non-trivial, steady-state solutions in 1.5 dimensions which are compactly supported.}  Should we make the ansatz that $B$ is identically zero for all times, we arrive at the \emph{relativistic Vlasov-Poisson system} (rVP):
\begin{equation}
\textrm{rVP:} \left\{ \begin{array}{l} \partial_t f + v(p)\cdot \nabla_q f + \sigma E\cdot\nabla_p f = 0\\
\nabla\cdot E = 4\pi (\rho - \overline{\rho})\\
\rho(t,q) = \int f(t,q,p)\;d^3p
 \end{array}\right. .
\end{equation}
In the case that the initial data, $f_0$, for rVM is spherically symmetric (and there is no stray electromagnetic radiation from sources at infinity), we obtain $B\equiv 0$ for all times without further ado.  Hence, we expect rVP to be significant for spherical, single-specie plasmas.

If we allow the parameter $\sigma$ appearing in rVP to become negative, we obtain a model which formally describes a gas of relativistic particles interacting through Newtonian gravitation.  Such a model might well be assumed to be valid for a sufficiently ``hot'' gas (so that the use of the relativistic velocity is justified) but rarefied enough that gravity is adequately modeled by the Poisson equation.  Currently, the only work along these lines known to the author is \cite{R94} wherein Rendall proves that sufficiently regular, asymptotically flat initial data for the fully covariant Vlasov-Einstein system launches solutions which are well approximated by the \emph{non-relativistic} Vlasov-Poisson system.  However in \cite{KTZ08}, Kiessling and Tahvildar-Zadeh proposed a novel scheme whereby this attractive version of rVP might result from a two-specie version of rVM wherein the oppositely charged species are distributed spherically.  The author has investigated this proposal, and the results will be reported elsewhere.

Since in the following we will examine both types of systems, we refer to rVP with the repulsive interaction as rVP$^+$ (or the \emph{plasma-physics case}) and rVP with the attractive potential as rVP$^-$ (or the \emph{astrophysical case}).  We will show in this paper that the behavior of all Fourier modes of linearized rVP$^{\pm}$ is decidedly different from that of the non-relativistic VP$^{\pm}$ --- not only the behavior of the superluminal modes as already noted by Schlickeiser.  Before we begin the discussion of linearized rVP$^{\pm}$, we close this introduction with a brief overview of rigorous results concerning the fully non-linear rVP systems.

One of the earliest papers to appear on rVP$^{\pm}$ is \cite{GS85} wherein Glassey and Schaeffer show that global classical solutions will exist for initial data that are spherically symmetric, compactly supported in momentum space, and vanish on characteristics with vanishing angular momentum which are in addition compactly supported in $\mathbb{R}^6$ and have $L^{\infty}$-norm below a critical constant $\mathcal{C}_{\infty}^{\pm}$, with $\mathcal{C}_{\infty}^{+} = \infty$ and $\mathcal{C}_{\infty}^{-} < \infty.$  More recently,  Kiessling and Tahvildar-Zadeh \cite{KTZ08} have extended the theorem of Glassey and Schaeffer for rVP$^-$ by proving global existence of classical solutions for initial data which satisfy the same basic requirements as above but are in $\mathfrak{P}_1\cap C^1$ \symbolfootnote[2]{ $\mathfrak{P}_n\cap C^k$ is the set of probability measures on $\mathbb{R}^6$ absolutely continuous w.r.t. Lebesgue measure whose first $n$ moments are finite and whose Radon-Nikodym derivative is $C^k$. } and have $L^{\beta}$-norm below a critical constant $\mathcal{C}_{\beta}^{-}$ with  $\mathcal{C}_{\beta}^{-} < \infty$, and $\mathcal{C}_{\beta}^{-}$ identically zero iff $\beta < 3/2.$  The authors explicitly computed $\mathcal{C}_{3/2}^{-}$ but characterized the constant for other values of $\beta > 3/2$ as a variational problem.  The constants for the remaining cases were computed by the author \cite{Y11-1} in terms of the famous Lane-Emden functions.

Glassey and Schaeffer also investigated what may happen when solutions to rVP$^-$ are launched by initial data with $\|f\|_{\infty} > \mathcal{C}_{\infty}^{-}$.  They proved that negative energy data lead to ``blow-up" (i.e. formation of a singularity) in finite time.  This is in sharp contradistinction to the non-relativistic Vlasov-Poisson system with attractive coupling (VP$^-$) which does not exhibit finite time blow-up for classical data.  Indeed, the possibility of collapse for solutions to rVP$^-$ is a primary motivation for studying the system - as the collapse is due solely to ``relativistic effects."  In \cite{LMR08b},  Lemou, M\'ehats, and Rapha\"el proved that systems launched by initial data with negative total energy approach a self-similar collapse profile. Around the same time, Kiessling and Tahvildar-Zadeh proved that any spherically symmetric classical solution of rVP$^-$ launched by initial data satisfying $f_0 \in \mathfrak{P}_3\cap C^1$ (along with other technical requirements) and having \emph{zero total energy} and total (scalar) virial less than or equal to $-1/2$ will blow up in finite time (Theorem 6.1 of \cite{KTZ08}).  However, they left open the question whether such initial data existed.  Explicit examples of such data were found by the author and reported in \cite{Y11-2}.

There has also been much work concerning the nonlinear stability of stationary solutions of rVP$^-$ and the dynamical details of the solutions which blow-up in finite time.  Had\v zi\'c and Rein \cite{HR07} showed the non-linear stability of a wide class of steady-state solutions of $\textrm{rVP}^-$ against certain allowable perturbations utilizing energy-Casimir functionals.  Shortly thereafter, Lemou, M\'ehats, and Rapha\"el \cite{LMR08a} investigated non-linear stability versus the formation of singularities in $\textrm{rVP}^-$ through concentration compactness techniques.

As for work on decay rates for the full rVP system in unbounded space, Horst \cite{H90} showed in 1990 that continuously differentiable, spherically symmetric initial data which are compactly supported launch solutions whose spatial matter distributions decay almost like $t^{-3}$ in $L^{\infty}$-norm (there is a logarithmic factor in the decay rate).  In 2009, Glassey, Pankavich, and Schaeffer \cite{GPS09} proved that non-trivial, continuously differentiable initial data for rVP$^{-}$ in 1.5 dimensions with compact support exhibit no decay whatsoever in $L^p$-norm for $p \in [1,\infty]$.  The results of this paper (though only at the level of the linearized equations) seem to suggest that this absence of rapid decay in $\rho$ persists in three-dimensions (though slower decay through dispersion is still to be expected).

The outline of the remainder of the paper is as follows.  Section \ref{results} provides a summary of our basic assumptions and results for the torus and the full space.  Following this, we collect all the relevant functional analytic details we shall need in Section \ref{Gev_sect}.  Sections \ref{torus_sect} and \ref{full_space_sect} provide the proofs of all theorems for the case of the torus and the full space, respectively.  Finally, we collect in the Appendix several important (but lengthy) calculations proving decay rates of certain functions appearing in sections \ref{torus_sect} and \ref{full_space_sect}.

\bigskip

\noindent \textbf{Acknowledgements:} The author wishes to thank Yves Elskens, Michael Kiessling, and Markus Kunze for numerous enlightening conversations and many helpful comments.  The author also wishes to thank an anonymous reviewer for pointing out a serious issue with the handling of the astrophysical case in the original formulation.

\section{Basic Setup and Statement of Results}\label{results}

In this section, we will give the basic setup for rVP on both the torus and the full space, list the basic assumptions on the equilibrium and initial data, and state the major results of the paper.  Despite the fact that the conclusions are similar (for a mode-by-mode study, at any rate), we have divided the discussion of the two cases into separate subsections.  The methods we use to attack the evolution of the Fourier modes in either case follow largely the treatment in \cite{GS94} for the linearized (non-relativistic) Vlasov-Poisson system.

\subsection{rVP on the Torus}

\subsubsection{Basic Setup}

The relativistic Vlasov-Poisson system on $\mathbb{T}_L^3 \times \mathbb{R}^3$ (the three-dimensional torus of volume $L^3$ in the spatial variables) is given by
\begin{equation}
\textrm{rVP}^{\pm}:\;\left\{ \begin{array}{r}\left(\partial_t + \frac{p}{\sqrt{1+\lvert p \rvert^2}} \cdot \nabla_q \pm \nabla_q\varphi_t(q) \cdot \nabla_p  \right)f_t(p,q)=0\\ \\ \triangle_q\varphi_t(q) = 4\pi \left(\int f_t(p,q)\; d^3p - ML^{-3}\right)  \end{array}\right.,
 \end{equation}
 where
 \begin{equation}
 M = \iint f_0(p,q)\; d^3pd^3q = \iint f_t(p,q)\; d^3pd^3q > 0.
 \end{equation}
 $\textrm{rVP}^{+}$ models a system with repulsive interaction (the \emph{plasma-physics} case) while $\textrm{rVP}^{-}$ models a system with attractive interaction (the \emph{astrophysical} case).  We will identify the torus as the cube $[-L/2,L/2]^3$ in $\mathbb{R}^3$ equipped with periodic boundary conditions.

Defining $\rho_t(q) = \int f_t(p,q)\;d^3p$, we can write $$\varphi_t(q) = (\triangle_q^{-1}\ast (\rho_t-ML^{-3}))(q)$$ where $\triangle_q^{-1}(q)$ is the fundamental solution to Laplace's Equation $$\triangle_q(\triangle_q^{-1}(q)) = 4\pi \left(\delta(q)-L^{-3}\right),$$ on the torus.  Hence, we obtain the equivalent integro-partial-differential equation:
 \begin{equation}
\partial_t f_t + \frac{p}{\sqrt{1+\lvert p \rvert^2}} \cdot \nabla_q f_t + \sigma \nabla_q(\triangle_q^{-1}\ast (\rho_t-ML^{-3})) \cdot \nabla_p f_t=0,\label{rVP}
 \end{equation}
where $\sigma = +1$ in the repulsive case and $\sigma = -1$ in the attractive case.

We wish to study the behavior of solutions to rVP$^\pm$ that are close (in some suitable sense) to a sufficiently nice steady-state solution $f_0(p) \ge 0$ (with total mass $L^3\int f_0 d^3p = M < \infty$).  Suppose that
\begin{eqnarray}
f_t(p,q) &=& f_0(p) + h_t(p,q),\\
\rho_t(q) &=& ML^{-3} + \int h_t(p,q) d^3p \equiv ML^{-3} + \rho^h_t(q),
\end{eqnarray}
with $h_t$ ``small'' compared to $f_0$ (at the very least, we will need $f_0 + h_t$ to remain non-negative).  We note that we can always assume the initial condition, $h_0$, is a \emph{neutral variation} of $f_0$ --- by which we mean
\begin{equation}
\iint h_0 d^3pd^3q = 0.
\end{equation}
If $h_0$ does not satisfy this, letting $m = \iint h_0 d^3pd^3q$, we can take $\tilde{h}_0 = h_0 - mf_0/M$ and $\tilde{f}_0 = (1 + m/M)f_0$ (which will still be positive as long as $h_0$ is a small variation of $f_0$).  Of course, this would necessitate replacing $M$ in the rVP system above by $M+m$, but this merely amounts to a redefinition of the total mass of the system.

Since the equilibrium steady state solution makes no contribution to the force term, we have
\begin{equation}
\partial_t h_t + \frac{p}{\sqrt{1+\lvert p \rvert^2}} \cdot \nabla_q h_t + \sigma \nabla_q(\triangle_q^{-1}\ast \rho^h_t) \cdot \nabla_p (f_0 + h_t)=0.
 \end{equation}
If $h_t$ is indeed a small perturbation of $f_0$, we can hope that the quadratic term in this integro-PDE makes little contribution to the dynamics.  Formally dropping this term, we arrive at the linearized relativistic Vlasov-Poisson equation:
\begin{equation}
\partial_t h_t + \frac{p}{\sqrt{1+\lvert p \rvert^2}} \cdot \nabla_q h_t + \sigma \nabla_q(\triangle_q^{-1}\ast \rho^h_t) \cdot \nabla_p f_0 =0. \label{linrVP}
 \end{equation}
We note that in both the full non-linear equation and the linearized equation for $h_t$, the neutral variation condition is propagated in time:
\begin{equation}
\iint h_t(p,q)d^3pd^3q = \iint h_0(p,q) d^3pd^3q = 0.
\end{equation}

We can give a formal solution to the linearized rVP system through Duhamel's principle.  Defining
\begin{eqnarray}
S_h(t,p,q) &=& \sigma \nabla_q(\triangle_q^{-1}\ast \rho^h_t)(q) \cdot \nabla_p f_0(p),\label{Sdef}\\
v(p) &=& \frac{p}{\sqrt{1+\lvert p \rvert^2}},\label{relvel}
\end{eqnarray}
we find that the solution to \eqref{linrVP} can be represented as
\begin{align}
h_t(p,q) = & h_0\left(p, q - v(p)t\right) \nonumber\\
& \;\;\;\;\;\;\;\;- \int_0^t S_h\left(\tau,p, q - v(p)(t-\tau)\right)d\tau.\label{Duhamel}
\end{align}

We wish to study the behavior of the spatial Fourier modes for this system.  Since we are on the torus, the wavevectors are discrete and can be indexed by $k\in \mathbb{Z}^3$.  The appropriate transform is
\begin{equation}
\hat{f}(k) = \frac{1}{L^3}\int_{\mathbb{T}_L^3} f(q) e^{-2\pi i \frac{k}{L} \cdot q}d^3q.
\end{equation}
We note that we are only interested in modes with $|k|>0$ as the zero mode (which is just the total mass of the system) is certainly conserved in time and equal to zero by our assumption that $h_0$ is a neutral variation.  Taking the transform of both sides of \eqref{Duhamel} we arrive at
\begin{align}
\hat{h}_t(p,k) = & \hat{h}_0(p,k)e^{-2\pi i \left(\frac{k}{L} \cdot v(p)\right)t}\\\nonumber
  & \;\;\;\;\;\;\;\;- \int_0^t \widehat{S_h}(\tau,p, k)e^{-2\pi i \left(\frac{k}{L} \cdot v(p)\right)(t-\tau)}d\tau.
\end{align}
Integrating in $p$, we see that
$$\widehat{\rho^h_t}(k) = \int \hat{h}_t(p,k) d^3p.$$  Since
 \begin{eqnarray*}
 \widehat{S_h}(t,p, k) &=&   \sigma 2\pi i \widehat{\triangle_q^{-1}}(k) \widehat{\rho^h_t}(k) \left(\frac{k}{L} \cdot \nabla_p \right) f_0(p)\\
 &=& -\frac{2i \sigma L^2}{|k|^2}   \widehat{\rho^h_t}(k) \left(\frac{k}{L} \cdot \nabla_p \right) f_0(p),
\end{eqnarray*}
we have the following equation for the Fourier modes:
\begin{equation}
\widehat{\rho^h_t}(k) = \alpha(k,t) + \int_0^t \beta(k,t-\tau)\widehat{\rho^h_{\tau}}(k)d\tau,\label{Volterra}
\end{equation}
where
\begin{eqnarray}
\alpha(k,t)&=& \int \hat{h}_0(p,k)e^{-2\pi i \left(\frac{k}{L} \cdot v(p)\right)t}d^3p, \label{alpha}\\
\beta(k,t) &=& \frac{2 i \sigma L}{|k|}  \int \left(\hat{k} \cdot\nabla_p\right) f_0(p)e^{-2\pi i \left(\frac{k}{L} \cdot v(p)\right)t}d^3p \label{beta},
\end{eqnarray}
and where $\hat{k}$ is the unit vector in the direction of $k$. Note that (just as in the non-relativistic case) the mode associated to the wavevector $k$ evolves independently of any other mode via a Volterra equation.

\subsubsection{Assumptions and Results}

First and foremost, we will easily deduce from \eqref{Volterra} the following:
\begin{tm}\label{no_exp_decay_torus}
For any kinetic equilibrium data, $f_0$, only depending on $|p|$ and initial data, $h_0$, symmetric enough that the rate of decay as $t$ tends to minus infinity matches the rate of decay as $t$ tends to infinity (in particular, if $h_0(p,-q)=\pm h_0(p,q)$),  exponential decay of Fourier modes is not possible for the linearized relativistic Vlasov-Poisson system on the torus.
\end{tm}
Hence, if we narrowly interpret Landau Damping as exponential decay of modes, then we see that this phenomenon cannot occur for rVP on the torus (at least under the typical symmetry assumptions invoked in most all results known about rVP).  Moreover, we will see that this theorem is a direct consequence of the universal speed limit imposed by relativity ($c=1$ in our units).  As such, this is likely to be true for any reasonable relativistic model of a plasma.

Despite this fact, it will be possible for modes to decay sub-exponentially (i.e. like $\exp\left(-|t|^{\delta}\right)$ for $0 < \delta < 1$).  To see this,  we make the following assumptions on the equilibrium, $f_0$, and initial datum, $h_0$:
\begin{asp}\label{asp 1}
Let $f_0$ be Schwartz class on $\mathbb{R}^3$ and $h_0$ be Schwartz class on $\mathbb{T}^3_L \times \mathbb{R}^3$.   Moreover, for a given, non-zero wavevector, $k$, suppose there is an $s_k>1$ and a constant $C_k>0$ so that
\begin{eqnarray}
\sup_{v \in B_1(0)} \left|\left(\hat{k}\cdot \nabla_v\right)^n\left(\frac{\hat{k}\cdot (\nabla_pf_0)(p(v))}{(1-|v|^2)^{5/2}}\right) \right| &\le& C_k^{n+1}(n!)^{s_k},\label{Gev_est_f_0}\\
\sup_{v \in B_1(0)} \left|\left(\hat{k}\cdot \nabla_v\right)^n\left(\frac{\hat{h}_0(p(v),k)}{(1-|v|^2)^{5/2}}\right) \right| &\le& C_k^{n+1}(n!)^{s_k},\label{Gev_est_h_0}
\end{eqnarray}
\noindent where $\hat{k}$ is the unit vector in the direction of $k$.  Also, assume $h_0(p,q)$ is such that $h_0(p,-q)=\pm h_0(p,q)$ and that $\widehat{h}_0(p,k) = \widehat{h}_0(|p|,k)$.   Finally, assume $f_0(p)$ is spherically symmetric and strictly decreasing in $|p|$.
\end{asp}
Here, we have used the relativistic formula for momentum in terms of velocity
\begin{equation}
p(v) = \frac{v}{\sqrt{1-|v|^2}},\label{rel_mom}
\end{equation}
which is just the inverse of \eqref{relvel}.

We should note that as $f_0$ is independent of $q$, the constants appearing in \eqref{Gev_est_f_0} could be chosen independent of $k$.  For our purposes, the estimates in the direction of the wavevector $k$ will be all that is required (even if $f_0$ admits a uniform estimate for these quantities).  Note that these assumptions are true for the most important case of the J\"uttner Distribution at temperature $T$:
\begin{equation}
f_{\textrm{J}}(|p|) = \frac{1}{4\pi k_B T K_2(1/k_B T)}e^{-\sqrt{1+|p|^2}/k_B T},\label{MJdist}
\end{equation}
where $k_B$ is Boltzmann's constant and $K_2$ is the modified Bessel function of the second kind with index 2.  This describes the thermodynamic equilibrium distribution of momenta in a spatially uniform, relativistic ideal gas (analogous to the Maxwellian Distribution for non-relativistic gases).  The exponent $s_k$ can be taken to be 3 for this distribution (as can be shown via standard facts about Gevrey class functions - c.f. \cite{R93} and Section \ref{Gev_sect} below).  We note that our results can be easily extended to the case where the equilibrium is decreasing and compactly supported (the critical constants below will need to change accordingly).

Given the assumptions above, we can show:
\begin{tm}\label{decay_rate_torus}
In the plasma-physics case ($\sigma = +1$), suppose
\begin{equation}
\left(\frac{|k|}{L}\right)^2  > 4\int_0^{\infty}|p|\left[2\arctanh(v(|p|))-v(|p|)\right]f_0(|p|)d|p| >0. \label{supcrit_p}
\end{equation}
In the astrophysical case ($\sigma = -1$), suppose
\begin{equation}
\left(\frac{|k|}{L}\right)^2  > 4\int_0^{\infty}\left[\sqrt{1+|p|^2} + \frac{|p|^2}{\sqrt{1+|p|^2}}\right]f_0(|p|)d|p| >0. \label{supcrit_g}
\end{equation}
If the equilibrium data and initial data meet the requirements of Assumption \ref{asp 1} for this $k$, then there exist positive constants $c_k$, $\epsilon_k$ so that for this particular wavevector
\begin{equation}
|\widehat{\rho^h_t}(k)| \le  c_k e^{-\epsilon_k t^{1/s_k}},
\end{equation}
for all $t>0$.
\end{tm}
\noindent We refer to modes whose wavevectors satisfy \eqref{supcrit_p} and \eqref{supcrit_g} as \emph{supercritical modes}.  The modes associated to the remaining (non-zero) wavevectors will naturally be called subcritical.  NB: Since the terms ``supercritical'' and ``subcritical" occur in a variety of different contexts with various meanings, we emphasize that our use of the terms is rather literal here.  Supercritical modes are those with wavevectors of magnitude strictly larger than the critical value (and so subcritical modes are those with wavevectors of magnitude less than or equal to the critical value).  Note that it is the supercritical modes which are damped out at a rather fast rate.  The behavior of the subcritical modes is more delicate, but in general we suspect that they will be damped out much slower than the supercritical modes (or perhaps not damped at all on the torus).  While we do not provide anything more than heuristic reasoning that the subcritical modes are not rapidly damped, our expectations seem to be well-founded.  For example, Lerche \cite{Le69} shows that subcritical modes (supra-luminous in his terminology) for the electric field are damped out only like $t^{-1}$ for a one-dimensional relativistic plasma excited by an initial impulse at the origin of infinitesimal duration.

\begin{figure}[ht]\centering
  \includegraphics[bb=70 200 546 585,clip=true,scale=0.65]{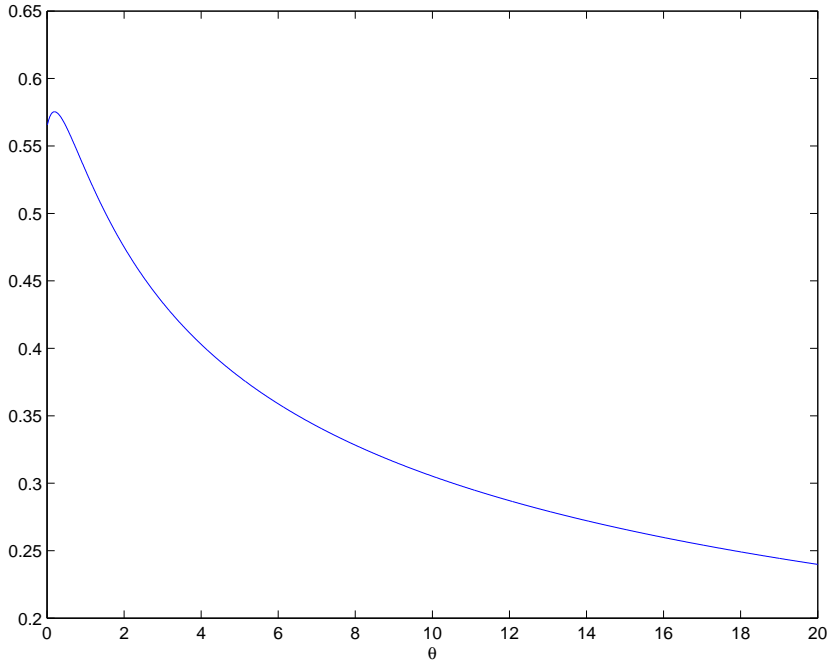}\\
  \caption{Supercritical Wavevector Size for Coulombic Plasmas as a function of $\theta = k_BT$ for the J\"uttner Distribution}
  \label{fig1}
\end{figure}

In Figure \ref{fig1}, we plot the square root of the right-hand side of \eqref{supcrit_p}  for the J\"uttner distribution \eqref{MJdist} as a function of $\theta = k_BT$.
Note that the maximum of this quantity occurs somewhere near $\theta = 0.2$ with a size of roughly $0.575$.  Since the zero mode is zero by our neutral variation assumption, the wavevector with the smallest possible non-zero magnitude for a given torus parameter, $L$, is of size $1/L$.  Hence, for tori of ``modest"\symbolfootnote[2]{Since we use units where $c=1$, taking time to be measured in seconds gives spatial units of light-seconds.  In these units, 0.05 light-seconds is roughly 15,000 km.  For comparison, the average equatorial diameter of the Earth is approximately 13,000 km (0.04 light-seconds) while the Sun has a diameter of roughly 1.4 million km (4.6 light-seconds).  The critical value of 1.7 light-seconds is roughly 510,000 km.} size, there may be no subcritical modes at all.  Indeed, for $0<L\lesssim 1.7$ there are no subcritical modes at any temperature for the J\"uttner distribution. Clearly, for a given temperature, $T,$ there is a critical torus dimension, $L_T,$ so that $L<L_T$ implies that all modes in the repulsive case are supercritical.

Next, we plot the square root of the right-hand side of \eqref{supcrit_g} for the J\"uttner distribution \eqref{MJdist} as a function of $\theta = k_BT$.
\begin{figure}[ht]\centering
  \includegraphics[bb=70 200 546 586,clip=true,scale=0.65]{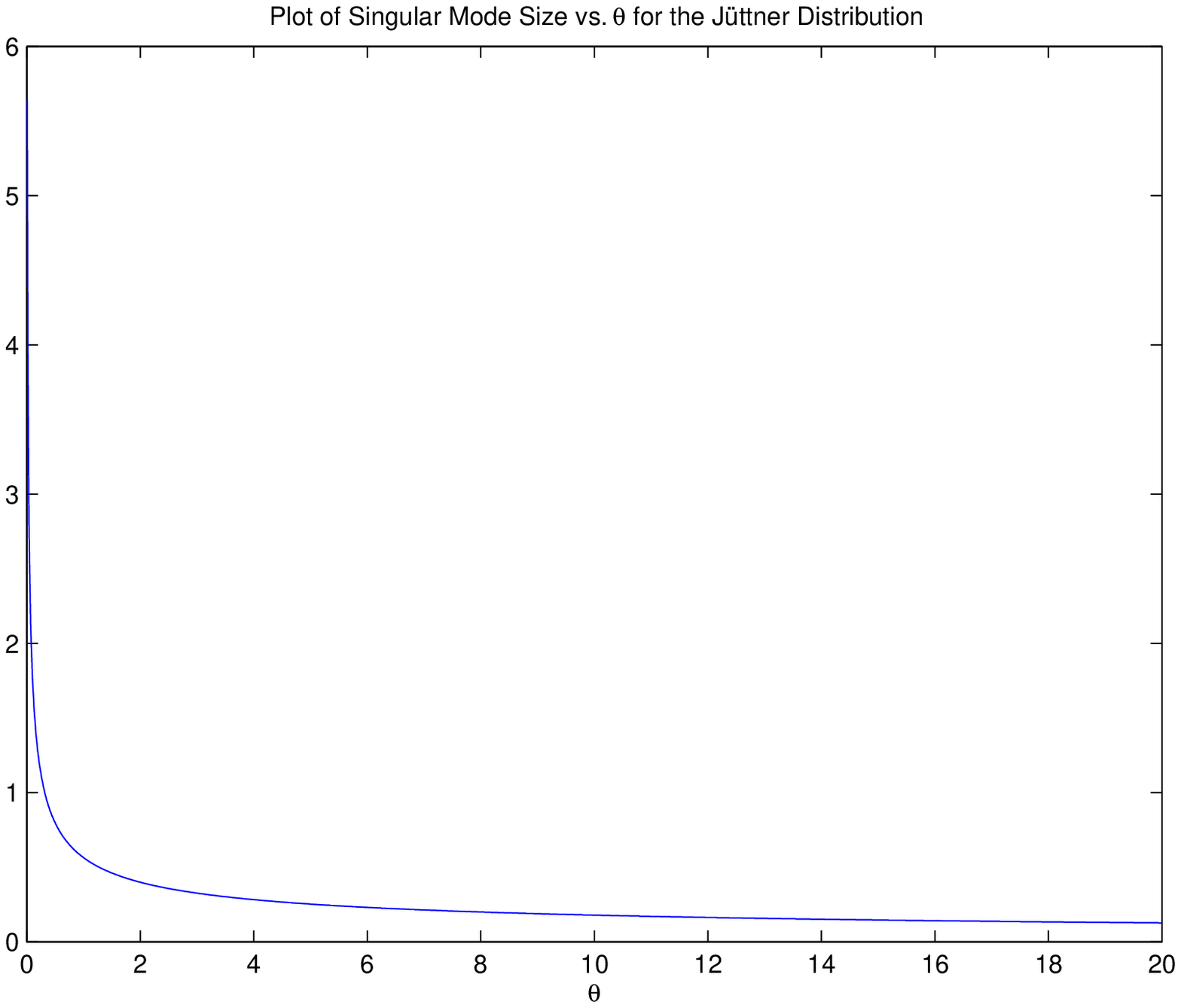}\\
  \caption{Supercritical Wavevector Size for Gravitational Plasmas as a function of $\theta = k_BT$ for the J\"uttner Distribution}
  \label{fig2}
\end{figure}
By way of comparison with the plasma-physics case, note that as the temperature of the gravitational plasma increases, the curve approaches zero.  So, for very hot gravitational plasmas (on the torus), it is possible that modes for every non-zero wavevector can be damped out.    As the temperature approaches zero, however, the curve diverges to positive infinity (numerical estimates show that it diverges like $\mathcal{C}/\sqrt{k_BT}$ for $\mathcal{C} \approx 0.564$).  Hence, colder gravitational plasmas can have relatively slow decay of modes associated to arbitrarily large wavevectors.

We should note that we have made no attempt to be optimal in our choice of Assumptions \ref{asp 1}.  These are merely the most convenient conditions which will ensure the rate of decay we wish to show is possible for relativistic plasmas.  Also, our technique does not easily address how the critical constant $\epsilon_k$ depends on the wavevector $k$.   Given that the wavevectors are discrete on the torus, we suspect that the constants $c_k, \epsilon_k,$ and $s_k$ can be chosen uniformly for all $k$ larger than a given magnitude (perhaps a magnitude slightly larger than the supercritical boundary).  Indeed, following the work of  Bedrossian, Masmoudi, and Mouhot \cite{BMM13} for non-relativistic plasmas, recent work by the author \cite{Y14} has shown that there can be uniform decay in $t$ for plasmas on the torus satisfying certain regularity conditions.  In particular, these conditions would require that there be no subcritical modes (other than $k=0$, naturally).  Of course, taking the torus parameter, $L$, sufficiently small will guarantee that all non-zero wavevectors are larger than a given magnitude.

\subsection{rVP on $\mathbb{R}^3$}

\subsubsection{Basic Setup}

Once we extend our examination to the entirety of $\mathbb{R}^3$, we are immediately forced to make a rather serious choice as to how to proceed.  As before, the Vlasov equation for rVP is given by:
\begin{equation}
\left(\partial_t + \frac{p}{\sqrt{1+\lvert p \rvert^2}} \cdot \nabla_q + \sigma \nabla_q\varphi_t(q) \cdot \nabla_p  \right)f_t(p,q)=0,
\end{equation}
where once again $\sigma = +1$ is the repulsive (plasma-physics) case while $\sigma = -1$ is the attractive (astrophysical) case.  The choice we must make is what sort of equilibrium solution to linearize about -- which will greatly influence exactly what the potential $\varphi_t(q)$ should be.

The most natural choice is to linearize about a sufficiently nice, time-independent solution, $f_0(p,q)$, of rVP (i.e. we consider solutions of the form $f_0(p,q) + h_t(p,q)$ for small $h_t$).  However, there are a number of issues which make an analysis in this case difficult.  If we write down a formal solution via Duhamel's principle, take the Fourier transform in the spatial variable, and integrate out the momentum variable, we find two major obstacles.  First,  the spatial transform will involve convolutions of the spatial transforms of  $\nabla_p h_t$ and $\nabla_q \varphi_t$ with transforms of analogous quantities coming from the equilibrium solution.  Hence, the time evolution of $\widehat{\rho_t^h}(\xi)$ will no longer evolve independently of the rest of the spectrum.  Second, integration of $\nabla_p h_t$ (or its transform in $q$) against $p$ does not give a quantity nicely related to $\rho_t^h$, and so we do not get a nice Volterra equation for $\widehat{\rho_t^h}(\xi)$.

The other option is to take a kinetic equilibrium, $f_0(p)$, which is uniform in the spatial variable.  This is, of course, much less reasonable from a physical viewpoint than the set-up above.  However, if the spatial support of the perturbation, $h_t$, is quite small compared to a natural length scale for a given inhomogeneous equilibrium $f_0(p,q)$ (or at least if the bulk of the perturbation lies in a compact set which is small on such a scale), then the fiction of a spatially homogeneous equilibrium is not entirely unreasonable.  However, we must assume that this spatially uniform equilibrium makes no contribution to the Poisson equation.  In the plasma-physics setting, this is usually accomplished by assuming there is a background neutralizing charge distribution (which is no more unreasonable than taking a uniform equilibrium $f_0$).  For the astrophysics case, this assumption is often referred to as the ``Jeans Swindle.''  As noted by Kiessling \cite{K03}, this assumption can be rigorously justified by an appropriate limiting procedure coupled with the fact that it is the forces (rather than the potential itself) which are important to the dynamics.

Given that such a maneuver is valid, if we linearize about a sufficiently nice, uniform kinetic equilibrium we obtain:
\begin{eqnarray}
\partial_t h_t + \frac{p}{\sqrt{1+\lvert p \rvert^2}} \cdot \nabla_q h_t + \sigma \nabla_q\varphi_t \cdot \nabla_p f_0 = 0,\label{Vlasov_homo_bg}\\
\triangle \varphi_t = 4\pi \int h_t(p,q) \;d^3p,
\end{eqnarray}
with the boundary condition that $\varphi_t(q) \asymp |q|^{-1}$ as $|q|\to \infty$.  This is essentially the same as what we found on the torus (and so many of the computations will be identical).

Formally, we can once again employ the construction used on the torus to ensure that \begin{equation} \iint h_0(p,q) \;d^3pd^3q = 0, \end{equation} (i.e. the neutral variation assumption).  However, the resultant $\tilde{h}_0= h_0 - \mathcal{C}f_0 $ will no longer decay nicely at infinity (unless, of course, $h_0$ does not decay at infinity  -- which stands in contradiction to the spirit of using the homogeneous equilibrium in the first place).  Hence, we will not make this assumption for $h_0$ in the whole space $\mathbb{R}^3$.  This will not be a serious impediment to the analysis going forward.  We should mention that the linearized flow is such that  $$\iint h_t(p,q) \;d^3pd^3q = \iint h_0(p,q) \;d^3pd^3q.$$

Once again, we can give a formal solution to the linearized rVP equation through Duhamel's principle:
\begin{align}
h_t(p,q) = & h_0\left(p, q - v(p)t\right) \nonumber\\
& \;\;\;\;\;\;\;\;- \int_0^t S_h\left(\tau,p, q - v(p)(t-\tau)\right)d\tau \label{Duhamel_rep},
\end{align}
where
\begin{equation}
S_h(t,p,q) = \sigma \nabla_q\varphi_t(q) \cdot \nabla_p f_0(p).\label{Sdef_H}
\end{equation}

For the Fourier transform, we take as our convention
\begin{equation}
\hat{f}(\xi) = \int f(q) e^{-2\pi i \xi \cdot q}d^3q.
\end{equation}
Taking the transform of both sides of \eqref{Duhamel_rep} we arrive at
\begin{align}
\hat{h}_t(p,\xi) = & \hat{h}_0(p,\xi)e^{-2\pi i \left(\xi \cdot v(p)\right)t}\\\nonumber
  & \;\;\;\;\;\;\;\;- \int_0^t \widehat{S_h}(\tau,p, \xi )e^{-2\pi i \left(\xi \cdot v(p)\right)(t-\tau)}d\tau.
\end{align}
Integrating in $p$, we see that
\begin{equation}
\widehat{\rho^h_t}(\xi) = \int \hat{h}_t(p,\xi) d^3p,
\end{equation}
 and
 \begin{equation}
 \widehat{S_h}(t,p, \xi) = -\frac{2i \sigma}{|\xi|}   \widehat{\rho^h_t}(k) \left(\hat{\xi} \cdot \nabla_p \right) f_0(p),
\end{equation}
where $\hat{\xi}$ is the unit vector in the direction of $\xi$.  Thus, we arrive at the following equation:
\begin{equation}
\widehat{\rho^h_t}(\xi) = \alpha(\xi,t) + \int_0^t \beta(\xi,t-\tau)\widehat{\rho^h_{\tau}}(\xi)d\tau,\label{Volterra2}
\end{equation}
where
\begin{eqnarray}
\alpha(\xi,t)&=& \int \hat{h}_0(p,\xi)e^{-2\pi i \left(\xi \cdot v(p)\right)t}d^3p, \label{alpha2}\\
\beta(\xi,t) &=& \frac{2 i \sigma}{|\xi|} \int \left(\hat{\xi}\cdot\nabla_p \right)f_0(p)e^{-2\pi i \left(\xi \cdot v(p)\right)t}d^3p \label{beta2}.
\end{eqnarray}
Note that as we take $\xi$ as a fixed parameter in our study going forward, \eqref{Volterra2} is formally no different than \eqref{Volterra} (all that is required is the replacement of $k/L$ by $\xi$).  Hence, much of the analysis for the full space will not differ significantly from that for the torus.

\subsubsection{Assumptions and Results}

As before, we will easily deduce from \eqref{Volterra2} the following:
\begin{tm}\label{no_exp_decay_R3}
For any kinetic equilibrium data, $f_0$, only depending on $|p|$ and initial data, $h_0$, symmetric enough that the rate of decay as $t$ tends to minus infinity matches the rate of decay as $t$ tends to infinity (in particular, if $h_0(p,-q)=\pm h_0(p,q)$),  exponential decay of Fourier modes is not possible for the linearized relativistic Vlasov-Poisson system on $\mathbb{R}^3$.
\end{tm}
The proof is more-or-less identical to that on the torus.  As such, we expect this to be true for any reasonable model of a relativistic plasma.

We make the following assumptions:
\begin{asp}\label{asp 2}
Suppose that $f_0$ is Schwartz class on $\mathbb{R}^3$ and $h_0$ is Schwartz class on $\mathbb{R}^6$.  Let $\xi \in \mathbb{R}^3$ (non-zero) be given, and suppose that there are constants $s_{\xi}>1$ and $C_{\xi}>0$ so that
\begin{eqnarray}
\sup_{v \in B_1(0)} \left|\left(\hat{\xi}\cdot \nabla_v\right)^n\left(\frac{\hat{\xi}\cdot (\nabla_pf_0)(p(v))}{(1-|v|^2)^{5/2}}\right) \right| &\le& C_{\xi}^{n+1}(n!)^{s_{\xi}},\label{Gev_est_f_0_2}\\
\sup_{v \in B_1(0)}  \left|\left(\hat{\xi}\cdot \nabla_v\right)^n\left(\frac{\hat{h}_0(p(v),\xi)}{(1-|v|^2)^{5/2}}\right) \right| &\le& C_{\xi}^{n+1}(n!)^{s_{\xi}},\label{Gev_est_h_0_2}
\end{eqnarray}
\noindent where $\hat{\xi}$ is the unit vector in the direction of $\xi$.  Moreover, assume $h_0(p,q)$ is such that $h_0(p,-q)=\pm h_0(p,q)$ and that $\widehat{h}_0(p,\xi) = \widehat{h}_0(|p|,\xi)$.   Finally, assume $f_0(p)$ is spherically symmetric and strictly decreasing in $|p|$.
\end{asp}
\noindent Once again, $p(v)$ is the relativistic momentum given by \eqref{rel_mom}.  As before, the assumptions on the equilibrium are more restrictive than necessary, but they include the J\"uttner Distribution \eqref{MJdist}.

Given these assumptions, we can show:
\begin{tm}\label{decay_rate_R3}
In the plasma-physics case ($\sigma = +1$), suppose
\begin{equation}
|\xi|^2  > 4\int_0^{\infty}|p|\left[2\arctanh(v(|p|))-v(|p|)\right]f_0(|p|)d|p| > 0. \label{supcrit2_p}
\end{equation}
In the astrophysical case ($\sigma = -1$), suppose that
\begin{equation}
|\xi|^2 > 4 \int_{0}^{\infty} \left(\sqrt{1+|p|^2}+\frac{|p|^2}{\sqrt{1+|p|^2}}\right)f_0(|p|)d|p| >0. \label{supcrit2_g}
\end{equation}
If the kinetic equilibrium and initial data satisfy the requirements of Assumption \ref{asp 2} for this particular $\xi$, then there exist positive constants $c_{\xi}$, $\epsilon_{\xi}$ so that
\begin{equation}
|\widehat{\rho^h_t}(\xi)| \le  c_{\xi} e^{-\epsilon_{\xi} t^{1/s_{\xi}}},
\end{equation}
for all $t>0$.
\end{tm}
\noindent We will once again refer to the wavevectors satisfying \eqref{supcrit2_p} and  \eqref{supcrit2_g} as supercritical.  As on the torus, the supercritical modes are the ones which can exhibit sub-exponential damping.  The behavior of the subcritical modes is far more delicate, and we still expect that they are damped out at a far slower rate.  Unlike the situation on the torus, there will always be subcritical modes, and it is the contribution of these modes which should dominate the large time behavior of the system.

As the method of proof is essentially the same, our technique does not easily address how $c_{\xi}, \epsilon_{\xi},$ and $s_{\xi}$ depend on $\xi$.  As we approach the supercritical boundary \eqref{supcrit_p} or \eqref{supcrit_g} (depending on the character of the interaction), $\epsilon_{\xi}$ should go to zero.  As above, this expectation comes from the results outlined in the introduction which indicate an overall rational rate of decay (at best) on $\mathbb{R}^3$.   Considerations in the appendix (where we collect several technical results) clearly show that we lose control over the estimates guaranteeing that our functions belong to a particular Gevrey class as we approach the critical boundary.\symbolfootnote[2]{Formally, we need control over the size of a function of the form $(1-f(\xi,\omega))^{-1}$ (where $\omega$ is dual to $t$ via Fourier Transform) uniform in $\omega$.  As $\xi$ approaches the critical boundary, $f(\xi,\omega)$ will approach $1$ for certain values of $\omega$ --- effectively destroying all estimates.}  Unlike the torus, we will definitely have wavevectors arbitrarily close to the critical boundary on the full space.  Hence, if it is at all possible to choose  $c_{\xi}, \epsilon_{\xi},$ and $s_{\xi}$ uniformly for $|\xi| >c$, then $c$ must likely be strictly larger than \eqref{supcrit_p} or \eqref{supcrit_g}.  The exact dependence of these important constants on the wavevector would certainly make for an interesting research project in the future.

\section{A Brief Overview of Gevrey Class Functions} \label{Gev_sect}

We give a brief resume of Gevrey functions and their properties which are of primary interest in this paper.  For more details, see Chapter 1 of \cite{R93}.  For an open subset, $\Omega$, of $\mathbb{R}^n$, a function is in the Gevrey class $G^s(\Omega)$ ($s \ge 1$) if $f$ is $\mathcal{C}^{\infty}(\Omega)$ and for every compact subset $K \subset \Omega$ there is a constant $C$ so that for every multi-index $\alpha$
\begin{eqnarray}
\left\|\partial^{\alpha}f \right\|_K &\le& C^{|\alpha|+1}(\alpha !)^s.
\end{eqnarray}
We should note that this is equivalent to the requirement that
 \begin{eqnarray}
\left\|\partial^{\alpha}f \right\|_K &\le& RC^{|\alpha|}(\alpha !)^s.
\end{eqnarray}
for some  positive constants $R$ and $C$ independent of $\alpha$ and $x \in K$. $G^s(\Omega)$ is a vector space closed under point-wise multiplication of functions and differentiation.  Clearly $G^s(\Omega) \subset G^t(\Omega)$ whenever $s \le t$.  The class $G^1(\Omega)$ corresponds exactly to the analytic functions in $\Omega$.

When $s>1$, there are functions in $$G^s_c(\Omega)\equiv \mathcal{C}^{\infty}_c(\Omega) \cap G^s(\Omega),$$ which are the compactly supported functions of Gevrey degree $s$ in $\Omega$.  Of course, $\mathcal{C}^{\infty}_c(\Omega) \cap G^1(\Omega) = \{0\}$ since no non-trivial analytic function can have compact support.  Via standard arguments, the spaces $G^s_c(\Omega)$ are dense in $\mathcal{C}^{\infty}_c(\Omega)$ for any $s>1$ (and so, dense in many other spaces).

The primary theorem we shall quote is the following (for a proof, see pp. 31--33 of \cite{R93}):
\begin{tm}[Fourier Transforms of Gevrey Functions]
\mbox{}\vspace{-\baselineskip}
\begin{enumerate}
\item[(i)] Assume $\varphi \in G^s_c(\mathbb{R}^n)$ ($s > 1)$; then there exist positive constants $C$ and $\epsilon$ so that
\begin{equation}
|\widehat{\varphi}(\xi)| \le C e^{-\epsilon |\xi|^{1/s}}.
\end{equation}

\item[(ii)] If the Fourier transform of $\varphi \in \mathcal{S}'(\mathbb{R}^n)$ satisfies the estimate above, then $\varphi \in G^s(\mathbb{R}^n)$.
\end{enumerate}
\end{tm}
\noindent $\mathcal{S}'(\mathbb{R}^n)$ above is the space of Tempered Distributions (dual to the Schwartz class, $\mathcal{S}(\mathbb{R}^n)$).

We will need a slight generalization of item (i) in the theorem above for dimension $n=1$.
\begin{cor}\label{Gevrey_cor}
Assume $\varphi \in G^s(\mathbb{R})$ for $s>1$ and that there exist a real number $\delta > 0$, an integer $m_0 \ge 0$, and a compact set $K \subset \mathbb{R}$ so that for all integers $m\ge0$
\begin{equation}
\int_{K^c}\left|\frac{d^m \varphi}{dx^m}\right|dx \le \delta^{m_0+m} (m_0+m)!,
\end{equation}
then there exist positive constants $C$ and $\epsilon >0$ so that
\begin{equation}
|\widehat{\varphi}(\xi)| \le C e^{-\epsilon |\xi|^{1/s}}.
\end{equation}
For $s=1$, the theorem holds but only with a rate of $1/s'$ for $s'>1$ in the exponential.
\end{cor}

\noindent \textbf{Proof:}  We follow the proof of item (i) in the theorem as given in \cite{R93}.  The key idea is that the decay rate above holds if and only if
\begin{equation}
|\xi|^{N/s}|\widehat{\varphi}(\xi)| \le C\left(CN\right)^N, \label{est}
\end{equation}
for all $N=1,2,3,\ldots$ and some constant $C$ independent of $N$.  Moreover, our assumption on the integrability of $\varphi$ ensures that $\widehat{\varphi}$ will be continuous (and vanishing at infinity) by the Riemann-Lebesgue lemma.  Hence, we need only verify \eqref{est} on $|\xi|>1$ since we can then ensure this estimate holds for all $\xi$ (by possibly making $C$ larger).  From hence forth, we assume $|\xi| > 1$.

We have
\begin{eqnarray}
\left|\xi^{m}\widehat{\varphi}(\xi)\right| &=& (2\pi)^{-m}\left|\widehat{\frac{d^m \varphi}{dx^m}}(\xi)\right|\\
&\le&(2\pi)^{-m}\left(\int_K \left|\frac{d^m\varphi}{dx^m}(x)\right|dx + \int_{K^c} \left|\frac{d^m\varphi}{dx^m}(x)\right|dx\right)\\
&\le& (2\pi)^{-m}\left(|K|C^{m+1}(m !)^s + \delta^{m_0+m} (m_0+m)!\right),
\end{eqnarray}
where the last inequality follows by the fact that $\varphi \in G^s(\mathbb{R})$ and the given estimate for $\varphi$ and its derivatives on $K^c$.  Since $m_0$ is fixed, for every $\gamma >0$, there is a constant $\mathcal{C} = \mathcal{C}(m_0,\gamma)$ so that $$(m_0+m)! \le \mathcal{C}(m!)^{1+\gamma}.  $$  Since $\delta$ is fixed and $s>1$, without loss of generality we can assume (by making $C$ larger if necessary)
\begin{equation}
\delta^{m_0+m} (m_0+m)!\le |K|C^{m+1}(m!)^s,
\end{equation}
for all $m\ge 0$ (note that this is where we use the fact that $s$ is strictly larger than 1). Thus,  we have
\begin{eqnarray}
|\xi|^{m}\left|\widehat{\varphi}(\xi)\right| &\le& 4\pi|K|\left(\frac{C}{2\pi}\right)^{m+1}(m !)^s\nonumber\\
&\le& 4\pi|K|\left(\frac{C}{2\pi}\right)^{m+1}(m)^{sm}.
\end{eqnarray}
Note that for the case $s=1$, we would need to use $(m!)^{1+\gamma}$ in the estimate above.  We may also assume that $C$ is greater than or equal to $2\pi$ (at the possible expense of making $C$ larger).

Let $M$ be the unique integer so that $N/s \le M< N/s+1$ and let $m = M$ in the inequality above.  Since we only need consider $|\xi| > 1$, we have
\begin{eqnarray}
|\xi|^{N/s}\left|\widehat{\varphi}(\xi)\right| &\le& |\xi|^{M}\left|\widehat{\varphi}(\xi)\right|\nonumber\\
&\le& 4\pi|K|\left(\frac{C}{2\pi}\right)^{M+1}(M)^{sM}\nonumber\\
&\le& \frac{|K|C^2}{\pi }\left(\frac{C}{2\pi}\right)^N(N+1)^{N+s},
\end{eqnarray}
where we have used the fact that $s\ge 1$ to surmise that $M < N+1$.  Note that this is where we need $C/2\pi \ge 1$.  Now, for some constant $\widetilde{C} = \widetilde{C}(s)>0$, we will have $$(N+1)^{N+s} \le \widetilde{C}^NN^N.  $$  Hence, we see
\begin{equation}
|\xi|^{N/s}\left|\widehat{\varphi}(\xi)\right| \le c\left(cN\right)^N,
\end{equation}
where $$ c=\max\left\{\frac{|K|C^2}{\pi }, \frac{C\widetilde{C}}{2\pi}\right\},  $$Hence, the corollary follows. $\;\;\;\; \square$

\noindent \textbf{Remark:}  This result should continue to hold true in $\mathbb{R}^n$.  Since we will only use this fact for the transform in the single variable $t$, the above corollary is sufficient for our purposes.

We will state a simple lemma which will make it easier to use Corollary \ref{Gevrey_cor}.
\begin{lem}\label{Gevrey_lem}
Suppose that there are functions $\varphi, \phi  \in \mathcal{C}^{\infty}(\mathbb{R})$ such that there exist real numbers $\delta, \epsilon > 0$, an integer $m_0 \ge 1$, and a compact set $K \subset \mathbb{R}$ so that for all integers $m\ge0$
\begin{eqnarray}
\left\|\frac{d^m \varphi}{dx^m}\right\|_{L^1(K^c)} &\le& \delta^{m_0+m} (m_0+m)!,\\
\left\|\frac{d^m \phi}{dx^m}\right\|_{L^{\infty}(K^c)} &\le&\epsilon^{m_0+m} (m_0+m)!.
\end{eqnarray}
Then there exists a constant $\mathcal{C}$ so that for all $m\ge 0$
\begin{equation}
\left\|\frac{d^m (\phi \cdot \varphi)}{dx^m}\right\|_{L^1(K^c)} \le \mathcal{C}^{m_0+m} (m_0+m)!.
\end{equation}
\end{lem}

\noindent \textbf{Proof:}  This result follows easily from the generalized Leibniz formula for derivatives of products:
\begin{eqnarray}
\int_{K^c}\left|\frac{d^m (\phi \cdot \varphi)}{dx^m}\right|dx &\le& \sum_{i=0}^m \binom{m}{i}\int_{K^c}\left|\frac{d^i \phi}{dx^i}\right|\left|\frac{d^{m-i} \varphi}{dx^{m-i}}\right|dx\nonumber\\
&\le& \sum_{i=0}^m \binom{m}{i} \left\|\frac{d^i \phi}{dx^i}\right\|_{L^{\infty}(K^c)}\left\|\frac{d^{m-i} \varphi}{dx^{m-i}}\right\|_{L^1(K^c)} \nonumber\\
&\le& \sum_{i=0}^m \binom{m}{i}\epsilon^{i+m_0} (i+m_0)!\delta^{m-i+m_0} (m-i+m_0)!.
\end{eqnarray}
Let $c = \max\{\delta, \epsilon \}$.  Then we have
\begin{equation}
\int_{K^c}\left|\frac{d^m (\phi \cdot \varphi)}{dx^m}\right|dx \le c^{n}c^{m_0+m}\sum_{i=0}^m \binom{m}{i} (i+m_0)! (m-i+m_0)!.
\end{equation}
Using the estimate $$(k+j)! \le 2^{k+j}k!j!,  $$  we see that
\begin{eqnarray}
\int_{K^c}\left|\frac{d^m (\phi \cdot \varphi)}{dx^m}\right|dx &\le& (m_0!)^2(2c)^{m_0}(2c)^{m_0+m} (m+1)!\nonumber\\
&\le&(m_0!)^2(2c)^{m_0}(2c)^{m_0+m} (m_0+m)!, \label{bad_est}
\end{eqnarray}
as $m_0 \ge 1$.  Since $m_0$ is fixed, \eqref{bad_est} implies that we can find a constant $\mathcal{C}$ satisfying the theorem.  $\square$

\noindent \textbf{Remarks:}  Once again, this result should continue to hold in $\mathbb{R}^n$.  Also the restriction that $m_0\ge 1$ is not too serious and can likely be relaxed with better estimates.  However, we will eventually use this result for functions which decay like $x^{-m_0}$ at infinity for some $m_0 \ge 1$ (and so this result will suffice).  Finally, by adapting the conditions on $\phi$ and $\varphi$ appropriately, H\"older's Inequality will give a similar result (with almost identical proof) for any pair of conjugate indices $p \ge 1$ and $q = p/(p-1)$.

\section{Linearized rVP on the Torus}\label{torus_sect}

\subsection{Evolution of the Spatial Fourier Modes}

For each fixed wavevector $k$, we have \eqref{Volterra} which is a Volterra equation for $\widehat{\rho^h_t}$.  Hence, the solution is best given in terms of the Laplace Transform:
\begin{equation}
\mathcal{L}[f](s) = \int_0^{\infty} f(t) e^{-st}dt,
\end{equation}
which is sensible at least for functions $f$ of exponential order at infinity and $s$ sufficiently large.  Formally taking the Laplace Transform of both sides of \eqref{Volterra} and solving for the transform of $\widehat{\rho^h_t}$ gives
\begin{equation}
\mathcal{L}[\widehat{\rho^h_{\cdot}}(k)](s) = \mathcal{L}[\alpha(k,\cdot)](s) + \frac{\mathcal{L}[\beta(k,\cdot)](s)}{1-\mathcal{L}[\beta(k,\cdot)](s)}\mathcal{L}[\alpha(k,\cdot)](s).
\end{equation}
Hence, we have
\begin{equation}
\widehat{\rho^h_t}(k) = \alpha(k,t) + \int_0^t \mathcal{I}(k,t-\tau)\alpha(k,\tau)d\tau, \label{soln}
\end{equation}
with the integral kernel $\mathcal{I}$ defined by
\begin{equation}
\mathcal{L}[\mathcal{I}(k,\cdot)](s)=\frac{\mathcal{L}[\beta(k,\cdot)](s)}{1-\mathcal{L}[\beta(k,\cdot)](s)} \label{kernel}
\end{equation}
whenever the inversion of the Laplace Transform is justified.  To make these formal calculations rigorous, we first examine basic properties of the functions $\alpha$ and $\beta$.  We will then need to determine exactly when taking the Laplace Transform of $\widehat{\rho}$ is justifiable.  Once that is settled, we show how our assumptions imply the sub-exponential rate of decay.

\subsection{Basic Properties of $\alpha(k,t)$ and $\beta(k,t)$}

First, note that both $\alpha$ and $\beta$ are of the form $$ g(k,t) = \int f(p,k)  e^{-2\pi i \left(\frac{k}{L} \cdot v(p)\right)t}d^3p.$$  Moreover, in both cases we expect $f(p,k)$ to decay in the variable $k$. For $\beta(k,t)$ we see that $$f(p,k) = \frac{2 i \sigma L}{|k|} \hat{k} \cdot  \nabla_p f_0(p),$$ and so $f$ decays like $|k|^{-1}$.  For $\alpha(k,t)$, $$f(p,k) =  \hat{h}_0(p,k)$$ (the spatial Fourier Transform of the initial data) which will decay to zero as $|k| \to \infty$ for any reasonable choice of initial data.

Our assumptions on $f_0$ and $h_0$ ensure that both will decay rapidly enough so that we may take their Fourier Transforms in $t$ (so long as $k \ne 0$).  For instance, taking the integral appearing in $\alpha$ and rewriting it in terms of $v = p/\sqrt{1+|p|^2}$ gives
\begin{equation}
\alpha(k,t) = \int_{B_1(0)} \frac{\widehat{h}_0\left(p(v),k\right)}{\left(1-|v|^2\right)^{5/2}}e^{-2\pi i \frac{kt}{L}\cdot v}d^3v,
\end{equation}
which explains the form of the estimates in our assumptions.  We know that this integral is well defined at $t=0$.  For $t >0$, we can integrate by parts in $v$ to transfer a directional derivative of the form $\hat{k}\cdot \nabla_v$ from the exponential over to the term involving $\widehat{h}_0$.  Each such integration by parts we perform adds a factor of $(|k|t)^{-1}$. As the resulting integrand is bounded (by assumption) and the integration is over the unit ball, we get an estimate on the decay rate.  Since we can do this as often as we like, we see that we have an estimate of the form
\begin{equation}
\left|\alpha(k,t)\right| \le \frac{d_{k,m}}{\left(1+|k|t\right)^m} \label{rationaldecay}
\end{equation}
for all $t>0$ and any $m \ge 1$.  The same estimates apply to $\beta$ (with a constant $d_m$ independent of $k$).  Hence, we can assume our generic function $g$ under consideration has sufficient decay in $t$ so that we may consider its Fourier Transform in this variable.

If we denote the Fourier Transform of $g(k,t)$ with respect to $t$  by $\widehat{g}^{t}(k,\omega)$ (to distinguish this transform from the one over the spatial variables), we have
\begin{align*}
\widehat{g}^{t}(k,\omega) &= \int f(p,k) \left[\int_{-\infty}^{\infty} e^{-2\pi i \left(\frac{k}{L} \cdot v(p)\right)t} e^{-2 \pi i \omega t}dt\right]d^3p\\
&=\int f(p,k) \; \delta\left(\omega +  \frac{k}{L} \cdot \frac{p}{\sqrt{1+\lvert p \rvert^2}}\right)d^3p,
\end{align*}
which is justifiable so long as $f$ is integrable over the (unbounded) surfaces $$\omega +  \frac{k}{L} \cdot \frac{p}{\sqrt{1+\lvert p \rvert^2}}=0$$ ($f$ being Schwartz class will certainly do the job).  Note that whenever this transform makes sense, $\widehat{g}^{t}(k,\omega)$ is compactly supported in the variable $\omega$ with support in the interval $|\omega| \le |k|/L$.  Hence, $g(k,t)$ will be $C^{\infty}$ as a function of $t$ whenever it is well-defined.  Of course, this can be seen directly from the definition of $g$!  If the function $f(p,k)$ makes sense in the integral defining $g$, then differentiating with respect to $t$ brings down a term of order $1$ in $p$.  Hence, dominated convergence gives existence of the derivative (and by induction, all higher derivatives).

Moreover, $g(k,t)$ will be analytic in $t$ (for fixed $k$) whenever it is well-defined.  Simply inserting the power-series representation for the exponential function into the definition of $g$ gives $$ g(k,t) = \sum_{n=0}^{\infty}\left[\int \left(\frac{k}{L} \cdot v(p)\right)^n f(p,k)\,  d^3p\right]\frac{(-2 \pi i t)^n}{n!}.$$  This form is justified since $$\left|\left(\frac{k}{L} \cdot v(p)\right)^n\right| \le \left(\frac{|k|}{L}\right)^n,$$ and $f(p,k)$ is an $L^1$ function of $p$ (for each fixed $k$).

Since the phenomenon of Landau Damping in plasmas is often construed as \emph{exponential} decay of Fourier Modes in $t$, our next question would be to determine if such decay is possible for the functions $\alpha$ and $\beta$.  Assumption \ref{asp 1} ensures that for the generic $g$ defined above any decay of $g$ as $t\to \infty$ is the same as the decay for $t\to -\infty$.

\bigskip

\noindent \textbf{Proof of Theorem \ref{no_exp_decay_torus}:}

Suppose that $g$ decays exponentially as $|t| \to \infty$.  Then the complex Fourier transform of $g$ in the variable $t$, $$\widetilde{g}^t(k,z) = \int g(k,t)e^{-2\pi i z t}dt,  $$ would be holomorphic in an infinite strip containing the real axis (c.f. Section IX.3 of \cite{RS75}).  However, we have already noted that on the real axis, this function is compactly supported!  Since we know the transform is not the zero function, this is impossible.  Hence, neither $\alpha$ nor $\beta$ can decay exponentially in time.  Likewise, taking the Fourier Transform in $t$ of \eqref{Volterra} shows that the transform of $\widehat{\rho^h_t}(k)$ will have compact support whenever the transform of $\alpha$ does.  Hence, $\widehat{\rho^h_t}(k)$ cannot decay exponentially in time either.  $\square$

\bigskip

This result is in sharp disagreement with the case of non-relativistic plasmas on the torus where exponential decay can and does occur (for sufficiently nice initial data).  Moreover, we see that the roadblock is precisely the fact that relativistic speeds are bounded above.  Thus, we expect this fact to hold true in fully covariant models (rather than being some artefact of the pseudo-relativistic nature of rVP). As we will detail below, while we cannot have exponential decay of Fourier modes in the relativistic case, we can have almost this much decay for nice enough initial data.

We should note that the theorem above does not entirely rule out exponential decay of modes.  The only way a given Fourier mode can exhibit exponential decay in the future is if the evolution in the past has a markedly slower rate of decay (at best sub-exponential).  In other words, exponential decay of Fourier modes can only occur when there is sufficient asymmetry in the evolution of the plasma.  Be sure to note that we make no guarantee of such decay even in this case!

\subsection{Proof of Theorem \ref{decay_rate_torus}}

In this section, we collect all the technical calculations which show that Assumption \ref{asp 1} implies the decay rate listed in Theorem \ref{decay_rate_torus}.  Proving this theorem entails a detailed examination of the integral kernel appearing in \eqref{Volterra} via the Laplace transform along with the results on Gevrey class functions from Section \ref{Gev_sect}.

\subsubsection{Decay Rates for $\alpha(k,t)$ and $\beta(k,t)$}

As we have seen above,  the evolution of the Fourier modes of the spatial distribution $\rho^h_t$ depend in a complicated way on the smoothness of the equilibrium $f_0$ and the initial datum $h_0$.  While we have ruled out the possibility of exponential decay for the Fourier modes, it is still possible that we can have sub-exponential decay with some exponent $0< \epsilon <1$: $$ |g(k,t)| \le \mathcal{C}_1e^{-\mathcal{C}_2|t|^{\epsilon}}.  $$  This brings us to the idea of Gevrey classes whose functions have exactly this property.

For the generic function $g(k,t)$ introduced in the previous section, we have already shown that $\widehat{g}^{t}(k,\omega)$ is compactly supported in $\omega$ for any given wavevector, $k$.  Thus, if we know that $\widehat{g}^{t}(k,\omega)$ is in $G^s(\mathbb{R})$ for a given $k$ and some $s>1$ (which potentially depends on $k$), then we can deduce that $g(k,t)$ decays like $|g(k,t)| \le C e^{-\epsilon |t|^{1/s}}$ for some constants $C$ and $\epsilon$ (also depending on $k$).  As such, we need to estimate the size of $\partial_{\omega}^n\widehat{g}^{t}$ over its support:
\begin{equation}
\textrm{supp}\left(\partial_{\omega}^n\widehat{g}^{t}\right)\subseteq \textrm{supp}\left(\widehat{g}^{t}\right) \subseteq \left\{\omega : |\omega| \le |k|/L\right\}.
\end{equation}
Switching the integration over to ``velocity space", we have
\begin{equation}
\widehat{g}^{t}(k,\omega) = \int_{B_1(0)} \frac{f(p(v),k)}{(1-|v|^2)^{5/2}} \delta\left(\omega +  \frac{k}{L} \cdot v \right)d^3v,
\end{equation}
where $B_1(0)$ is the ball of radius $1$ centered at the origin, and $$ p(v) = \frac{v}{\sqrt{1-|v|^2}}. $$  A short calculation gives
\begin{align}
\partial_{\omega}^n\widehat{g}^{t}(k,\omega) = (-1)^n\frac{L^{n}}{|k|^{n}}\int_{B_1(0)} \left(\hat{k}\cdot \nabla_v\right)^n&\left(\frac{f(p(v),k)}{(1-|v|^2)^{5/2}}\right) \nonumber\\ &\;\;\;\;\cdot\delta\left(\omega +  \frac{k}{L} \cdot v \right)d^3v,
\end{align}
where $\hat{k}$ is the unit vector in the direction of $k$ (our assumptions in momentum space ensure that $f$ and its derivatives will vanish fast enough at the boundary of the unit sphere to make the integrals above sensible).

From the above calculation, we see that if
\begin{equation}
\left|\left(\hat{k}\cdot \nabla_v\right)^n\left(\frac{f(p(v),k)}{(1-|v|^2)^{5/2}}\right) \right| \le C_k^{n+1}(n!)^s,\label{Gev_est_f}
\end{equation}
for all $v\in B_1(0)$ and some $s>1$, then
\begin{eqnarray}
\left|\partial_{\omega}^n\widehat{g}^{t}(k,\omega) \right| &\le& C_k\left(\frac{C_kL}{|k|}\right)^n(n!)^s \int_{B_1(0)} \delta\left(\omega +  \frac{k}{L} \cdot v \right) d^3v\\
&\le& \pi\left(\frac{C_kL}{|k|}\right)^{n+1}(n!)^s.
\end{eqnarray}
The second inequality follows from the first as the remaining integral is merely the surface area of the portion of the plane $ k/L\cdot v = -\omega$ which intersects the unit ball (and so is bounded by $\pi$) multiplied by a factor of $L/|k|$ coming from rescaling the Dirac delta.  Therefore, if the generic function $f$ satisfies \eqref{Gev_est_f} for a given $k$, then $\widehat{g}^{t}$ will be in $G^s_c(\mathbb{R})$, and we will have sub-exponential decay in $t$ for this particular wavevector $k$.

Recall that for $\alpha(k,t)$, the appropriate function to consider is $$f(p(v),k) = \hat{h}_0(p(v),k),  $$ while for $\beta(k,t)$ we have $$f(p(v),k)=\frac{2i\sigma L}{|k|}\hat{k}\cdot(\nabla_pf_0)(p(v)).  $$  From our calculations above, we see that Assumption \ref{asp 1} implies that there are positive constants $c_k$ and $\epsilon_k$ so that
\begin{equation}
|\alpha(k,t) | , |\beta(k,t) | \le c_k e^{-\epsilon_k |t|^{1/s_k}},\label{a b decay}
\end{equation}
which greatly improves our simpler (rational) decay rates deduced earlier.  Moreover, by our earlier computations, we know that these functions are in $G^1(\mathbb{R})$ as a function of $t$, and by the above decay rate, we see that in fact
\begin{equation}
\alpha(k,\cdot), \beta(k,\cdot) \in G^1(\mathbb{R})\cap L^p(\mathbb{R}),
\end{equation}
for all $p\ge 1$.

Note that we have not (as yet) made explicit use of the symmetry (or monotonicity) of $f_0$ or the assumption on the $p$ dependence of $\widehat{h}_0$.  These will play a role in the subsequent analysis of the convolution integral involving the kernel $\mathcal{I}$ (and the function $\alpha$).

\subsubsection{Existence of the Transform}

In our formal manipulations which led to the definitions of the integral kernel, $\mathcal{I}$, we have tacitly assumed that the transforms in question exist.  To tackle the question of exactly when this is justifiable, we begin with a preliminary analysis inspired by \cite{GS94}[\S 2].  We define
\begin{equation}
\left\|\widehat{\rho^h_t}(k)\right\|_1 \equiv \sup_{0\le \tau \le t}(1+\tau)\left|\widehat{\rho^h_{\tau}}(k)\right|.
\end{equation}
Using this norm with \eqref{Volterra} gives
\begin{eqnarray}
\left|\widehat{\rho^h_t}(k) \right| &=& \left|\alpha(k,t)\right| + \left\|\widehat{\rho^h_t}(k)\right\|_1 \int_0^t \left|\beta(k,t-\tau)\right|(1+\tau)^{-1}d\tau\\
&\le&c_k e^{-\epsilon_k t^{1/s_k}} + d_2\left\|\widehat{\rho^h_t}(k)\right\|_1 \int_0^t \frac{d\tau}{(1+\tau)\left(1+|k|(t-\tau)\right)^2},
\end{eqnarray}
where we have used the sub-exponential decay shown above for $\alpha$.  For $\beta$, we have used \eqref{rationaldecay} with $m=2$ since it has the advantage of having an explicit $|k|$ dependence (as opposed to the sub-exponential decay whose constants are a little more abstract).  Using basic estimates (and $|k| \ge 1$), we find for $t\gg 1$
\begin{equation}
\left|\widehat{\rho^h_t}(k) \right| \le c_k e^{-\epsilon_k t^{1/s_k}} + \frac{d_2}{|k|} \frac{\mathcal{C}}{1+t}\left\|\widehat{\rho^h_t}(k)\right\|_1
\end{equation}
for some constant $\mathcal{C}$ independent of $f_0$ and $g_0$.  Taking the norm of both sides yields
\begin{equation}
\left\|\widehat{\rho^h_t}(k) \right\|_1 \le \mathcal{C}_1  + \frac{\mathcal{C}_2d_2}{|k|}\left\|\widehat{\rho^h_t}(k)\right\|_1
\end{equation}
where $\mathcal{C}_1=\mathcal{C}_1(c_k, s_k)$ and $\mathcal{C}_2$ is a universal constant.  Hence, we get an estimate of the form
\begin{equation}
\left|\widehat{\rho^h_t}(k) \right| \le \frac{\mathcal{C}_3(c_k, \epsilon_k, s_k)}{1+t}
\end{equation}
(and so $\widehat{\rho^h_t}$ is an $L_2$-function of $t$) provided $\mathcal{C}_2d_2 < |k|$.  So, we see that taking the Fourier Transform of $\widehat{\rho^h_t}$ in $t$ is certainly justifiable for large enough $|k|$.  To get a more precise handle on exactly how big is big enough, we appeal to the Paley-Wiener Theory for Volterra Equations (c.f. \cite{GLS90}[Chap. 2]).

In a nutshell, Paley-Wiener Theory considers generic fixed point problems of the form $$u(t) = \phi(t) + k\ast u (t),  $$ for $k \in L^1_t$, $k(t)\le e^{Ct}$ ($C>0$), and $\phi \in L^1_{loc}(\mathbb{R}_+)$ where we extend $u$ and $\phi$ to be zero on $\mathbb{R}_-$ (so that the convolution is sensible).  The solution to this problem is $u = \phi-\phi \ast r$ where $r\in L^1_{loc}(\mathbb{R}_+)$ is the unique solution to $r = k + k \ast r .$  The major result of this theory is that $r \in L^1(\mathbb{R}_+)$ iff the Fourier-Laplace Transform of $k$ is never equal to 1 on the right half-plane $\Re z \ge 0$ (depending, or course, on your conventions for the transform) .  Once $r$ is in $L^1$, if we then know $\phi$ is also in $L^1$, we find that $u \in L^1(\mathbb{R}_+)$.  Once we know a function is $L^1$, we can certainly take its transform.

Certainly $\alpha$ and $\beta$ meet all the bounds and integrability requirements of $\phi$ and $k$, respectively.  The only thing to explore is the Fourier-Laplace Transform (in $t$) of $\beta$ on the right half-plane.  Our convention for this transform will be
\begin{eqnarray}
\mathcal{L}[f](k,z) &=& \int_0^{\infty}f(k,t)e^{-zt}dt\\
&=& \int_{-\infty}^{\infty}H(t)f(k,t)e^{-zt}dt,
\end{eqnarray}
where $H(t)$ is the unit Heaviside function.  If we write $z = x + 2\pi i y$ then we see that
\begin{eqnarray}
\mathcal{L}[f](k,x + 2\pi i y) &=& \int_0^{\infty}f(k,t)e^{-xt}e^{-2\pi i y t}dt \nonumber \\
&=& \mathcal{F}\left\{H \exp(-x\cdot)\right\}\ast \widehat{f}(y).
\end{eqnarray}
As is well known,
\begin{equation}
\mathcal{F}\left\{H \exp(-x\cdot)\right\}(y) = \left\{\begin{array}{cl} \left(x+2\pi i y\right)^{-1}, &  x>0\\ \frac{1}{2}\left(\frac{\textrm{P.V.}}{i \pi y} + \delta(y)\right), & x = 0 \end{array} \right.
\end{equation}
where the last expression is understood in the sense of distributions.

To begin,  a quick calculation shows that under the assumption that $f_0$ be spherically symmetric
\begin{equation}
\widehat{\beta}^t(k,y) = \left\{ \begin{array}{rc}\frac{4 \pi i \sigma L^3}{|k|^3} y \int_{\mathcal{P}(|y|L/|k|)}^{\infty}(1+|p|^2)  (-f_0'(|p|))d|p|, & |y| < \frac{|k|}{L}\\ 0, & |y|\ge \frac{|k|}{L}\end{array}\right.,\label{beta_trans}
\end{equation}
where we have defined $$\mathcal{P}(x) = \frac{x}{\sqrt{1-x^2}}.  $$  Note that $\widehat{\beta}^t$ is an odd function of $y$.   Also note that under our assumption $f_0$ be strictly decreasing, $\widehat{\beta}^t$ is strictly non-zero on the interior of its support with the obvious exception of $y=0$.

\textbf{Case 1: $x >0$.}  Here we have
\begin{align}
\mathcal{L}[\beta]&(k,x + 2\pi i y) = \int \left(x+2\pi i (y-\tau)\right)^{-1}\widehat{\beta}^t(k,\tau)d\tau \nonumber\\
&= \frac{4 \pi i \sigma L^3}{|k|^3} \int_{-\frac{|k|}{L}}^{\frac{|k|}{L}}\int_{\mathcal{P}(|\tau|L/|k|)}^{\infty} \frac{\tau}{x+2\pi i (y-\tau)}(1+|p|^2)  (-f_0'(|p|))d|p|d\tau \nonumber\\
&= \frac{4 \pi i \sigma L^3}{|k|^3} \int_0^{\infty} \left[\int_{-\frac{|k|}{L}v(|p|)}^{\frac{|k|}{L}v(|p|)}\frac{\tau}{x+2\pi i (y-\tau)}d\tau\right](1+|p|^2)  (-f_0'(|p|))d|p|\nonumber\\
&= \frac{4  \sigma L^2}{|k|^2} \int_0^{\infty}\left[ \frac{x+2\pi i y}{2\pi i |k|/L}\arctanh\left(\frac{2\pi i |k|/L}{x+2\pi i y}v(|p|)\right)-v(|p|)\right](1+|p|^2)  (-f_0'(|p|))d|p|
\end{align}
where $v(|p|) = |p|/\sqrt{1+|p|^2}$ is the inverse of $\mathcal{P}.$  Note that since $x\ne0$, the argument of the $\arctanh$ is always in a sensible range for this function.

The imaginary part of the integrand is controlled by
\begin{align}
\Im & \left[z\arctanh\left(\frac{v}{z}\right)\right] \nonumber\\
&=(\Im z)\ln\left(\frac{|z+v|}{|z-v|}\right) + (\Re z)\left[\arctan\left(\frac{\Im z}{\Re z +v}\right) - \arctan\left(\frac{\Im z}{\Re z -v}\right)\right]
\end{align}
where $v \in [0,1]$.   For our purposes, $z = \frac{L}{|k|}\left(y - i\frac{x}{2\pi}\right)$ with $x >0$.  Hence, the imaginary part of the integral comes from an integral involving
\begin{align}
-\left(\frac{Lx}{4\pi |k|}\right)&\ln\left(\frac{\frac{x^2}{4\pi^2} + \left(y+\frac{v|k|}{L}\right)^2}{\frac{x^2}{4\pi^2} + \left(y-\frac{v|k|}{L}\right)^2}\right) \nonumber \\
& + \left(\frac{Ly}{|k|} \right)\left[\arctan\left(\frac{x}{2\pi \left(y -\frac{|k|v}{L}\right)}\right) - \arctan\left(\frac{x}{2\pi \left(y+\frac{|k|v}{L}\right)}\right)\right]
\end{align}
Note that this quantity is zero when $v=0$.  Taking the derivative of this quantity with respect to $v$ yields
\begin{equation}
\frac{2xyv^2|k|^2}{\pi L^2}\left(\frac{x^2}{4\pi^2}+\left(y-\frac{v|k|}{L}\right)^2\right)^{-1}\left(\frac{x^2}{4\pi^2}+\left(y+\frac{v|k|}{L}\right)^2\right)^{-1}
\end{equation}
which is positive on $x>0, y>0$ and negative on $x>0, y<0$.  Hence, the Fourier-Laplace transform cannot equal $1$ in the region $x>0$ unless $y=0$.

If we consider the real axis, we have
\begin{equation}
\mathcal{L}[\beta](k,x)\!\! = \!\! \frac{-4\sigma L^2}{|k|^2}\int_0^{\infty}\!\! \left[v(|p|) - \frac{Lx}{2\pi |k|}\arctan\left(\frac{2\pi |k|}{Lx}v(|p|)\right)\right](1+|p|^2)  (-f_0'(|p|))d|p|.
\end{equation}
The quantity in square brackets is easily shown to be non-negative over the range of integration and decreasing in $x$.  Thus for $\sigma = +1$ (the plasma-physics case), the Fourier-Laplace Transform cannot equal $1$ in the region $x>0$.  For $\sigma = -1$ (the gravitational case), we can ensure that the Fourier-Laplace transform is never equal to one if we insist that
\begin{eqnarray}
\lim_{x \to 0} \mathcal{L}[\beta](k,x) &=& \frac{-4\sigma L^2}{|k|^2}\int_0^{\infty}v(|p|)(1+|p|^2)  (-f_0'(|p|))d|p| \nonumber\\
&=& \frac{-4\sigma L^2}{|k|^2}\int_0^{\infty}\left(\sqrt{1+|p|^2} + \frac{|p|^2}{\sqrt{1+|p|^2}}\right)  f_0(|p|) d|p| < 1.
\end{eqnarray}
Thus, we must require
\begin{equation}
\frac{|k|^2}{L^2} > 4 \int_0^{\infty}\left(\sqrt{1+|p|^2} + \frac{|p|^2}{\sqrt{1+|p|^2}}\right)  f_0(|p|) d|p|
\end{equation}
for gravitational plasmas.  Not only will we see that this matches up with the expression for the transform on the imaginary axis ($x=0$), but we will find that there is no other place that the transform can be equal to $1$ for the gravitational case.

\textbf{Case 2: $x =0$.}  Here we have
\begin{align}
\mathcal{L}[\beta]&(k,2\pi i y) = \frac{1}{2}\textrm{P.V.}\int \frac{\widehat{\beta}^t(k,y-\tau)}{i \pi \tau}d\tau  + \frac{1}{2}\widehat{\beta}^t(k,y)\nonumber\\
&=\frac{1}{2}\widehat{\beta}^t(k,y) - \frac{i}{2\pi}\int_{0}^{\infty}\frac{\widehat{\beta}^t(k,y-\tau)-\widehat{\beta}^t(k,y+\tau)}{\tau}d\tau.
\end{align}
Since $\widehat{\beta}^t(k,y)$ is purely imaginary (by the fact the background is spherical), we see that
 \begin{eqnarray}
 \Re\left\{\mathcal{L}[\beta](k,2\pi i y)\right\} &=& - \frac{i}{2\pi}\int_{0}^{\infty}\frac{\widehat{\beta}^t(k,y-\tau)-\widehat{\beta}^t(k,y+\tau)}{\tau}d\tau,\\
 \Im\left\{\mathcal{L}[\beta](k,2\pi i y)\right\} &=& -\frac{i}{2}\widehat{\beta}^t(k,y).
 \end{eqnarray}
We note that the imaginary part of $\mathcal{L}[\beta](k,2\pi i y)$ is compactly supported in $|y| \le |k|/L$ for each fixed wavevector $k$.

 Given the form of $\widehat{\beta}^t$ above \eqref{beta_trans} and our assumption that $f_0$ be strictly decreasing, we see that the imaginary part of  $\mathcal{L}[\beta](k,2\pi i y)$ is strictly non-zero on the interior of its support with the obvious exception of $y=0$.  Thus, $\mathcal{L}[\beta](k, 2\pi i y)$ cannot be equal to one on the set $|y| < \frac{|k|}{L}$ except possibly at $y=0$, and so any other potential singularities of $\mathcal{L}[\mathcal{I}(k,\cdot)](2\pi i y)$ must lie in $|y| \ge \frac{|k|}{L}$.

For the case $y=0$, we have
 \begin{eqnarray}
 \mathcal{L}[\beta](k,0) &=& \Re\left\{\mathcal{L}[\beta](k,0)\right\}\nonumber\\
 &=&  \frac{i}{\pi}\int_{0}^{\infty}\frac{\widehat{\beta}^t(k,\tau)}{\tau}d\tau\nonumber\\
 &=& -\frac{4\sigma L^3}{|k|^3}\int_{0}^{|k|/L}\int_{\mathcal{P}(\tau L /|k|)}^{\infty}(1+|p|^2)(-f_0'(|p|))d|p|d\tau\nonumber\\
 &=& -\frac{4\sigma L^2}{|k|^2} \int_{0}^{\infty} |p|\sqrt{1+|p|^2}(-f_0'(|p|))d|p|\nonumber\\
 &=& -\frac{4\sigma L^2}{|k|^2} \int_{0}^{\infty} \left(\sqrt{1+|p|^2}+\frac{|p|^2}{\sqrt{1+|p|^2}}\right)f_0(|p|)d|p|,
\end{eqnarray}
which yields the same conclusions as our considerations from the case $x>0$.  Hence, there is nothing new here.

We are left with $|y| \ge \frac{|k|}{L}$. Since we also have $\tau \ge 0$, $\widehat{\beta}^t(k,y+\tau)$ is identically zero on $y \ge \frac{|k|}{L}$, while $\widehat{\beta}^t(k,y-\tau)$ is identically zero for  $y \le -\frac{|k|}{L}$.  Thus, for $|y| \ge \frac{|k|}{L}$ we have
\begin{eqnarray}
 \mathcal{L}[\beta](k, 2\pi i y) &=& \Re \left\{\mathcal{L}[\beta](k, 2\pi i y)\right\} \nonumber\\
 &=& \left\{\begin{array}{cc}- \frac{i}{2\pi}\int_{y-\frac{|k|}{L}}^{y+\frac{|k|}{L}}\frac{\widehat{\beta}^t(k,y-\tau)}{\tau}d\tau & y \ge \frac{|k|}{L} \\ \frac{i}{2\pi}\int_{-y-\frac{|k|}{L}}^{-y+\frac{|k|}{L}}\frac{\widehat{\beta}^t(k,y+\tau)}{\tau}d\tau & y \le -\frac{|k|}{L}\end{array}\right.\\
 &=& - \frac{i}{2\pi}\int_{|y|-\frac{|k|}{L}}^{|y|+\frac{|k|}{L}}\frac{\widehat{\beta}^t(k,|y|-\tau)}{\tau}d\tau,
 \end{eqnarray}
 where for the last equality we have used the fact that $\widehat{\beta}^t$ is an odd function of its second variable.  As this is clearly even in $y$, we need only consider  $y \ge |k|/L$.

 Restricting ourselves to $y \ge |k|/L$, we have
  \begin{eqnarray}
 \mathcal{L}[\beta](k, 2\pi i y) &=&  - \frac{i}{2\pi}\int_{y-\frac{|k|}{L}}^{y+\frac{|k|}{L}}\frac{\widehat{\beta}^t(k,y-\tau)}{\tau}d\tau \nonumber\\
 &=& \frac{2\sigma L^3}{|k|^3}\int_{y-\frac{|k|}{L}}^{y+\frac{|k|}{L}} \frac{y-\tau}{\tau} \int_{\mathcal{P}(|y-\tau|L/|k|)}^{\infty}\!\!\!\!\!\!\!\!(1+|p|^2)  (-f_0'(|p|))d|p|d\tau\nonumber\\
 &=& \frac{4\sigma L^2}{|k|^2}\int_0^{\infty}\left[ \frac{yL}{|k|}\arctanh\left(\frac{|k|}{yL}v(|p|)\right)\!-\!v(|p|)\right]\nonumber\\
 &&\;\;\;\;\;\;\;\;\;\;\;\;\;\;\;\;\;\;\;\;\;\;\;\;\;\;\;\;\;\;\;\;\;\;\;\;\;\; \cdot (1+|p|^2)  (-f_0'(|p|))d|p| \label{Lbetaform}
 \end{eqnarray}

The function
\begin{equation}
F(x,v) = x \; \arctanh \left(\frac{v}{x} \right) - v,
\end{equation}
appearing in the integrand above can be shown to be non-negative and decreasing in $x$ over our range of interest ($x \ge 1$ and $0 \le v < 1$).   For more details, see the appendix at the end of the paper where this function is studied in great detail.  These results imply that $\mathcal{L}[\beta](k, 2\pi i y)$ is a decreasing function of $y$ for $y \ge |k|/L$ in the plasma-physics case ($\sigma = +1$).  So, should $\mathcal{L}[\beta](k, 2\pi i |k|/L)$ be less than one in this case, $\mathcal{L}[\beta]$ cannot equal to $1$ on the imaginary axis.  In the astrophysical case ($\sigma = -1$), $\mathcal{L}[\beta](k, 2\pi i y)$ is strictly negative for  $y \ge |k|/L$ and so cannot equal $1$.

At $y = |k|/L$, we have in the repulsive case
 \begin{align}
 \mathcal{L}[\beta](k, 2\pi i |k|/L) &= \frac{4L^2}{|k|^2}\int_0^{\infty}\left[\arctanh\left(v(|p|)\right)\!-\!v(|p|)\right]\nonumber\\
 &\;\;\;\;\;\;\;\;\;\;\;\;\;\;\;\;\;\;\;\;\;\;\;\;\;\;\;\;\;\;\;\;\;\;\;\;\;\; \cdot (1+|p|^2)  (-f_0'(|p|))d|p|\\
 &=\frac{4L^2}{|k|^2}\int_0^{\infty}\!\!|p|\left[2\arctanh(v(|p|))-v(|p|)\right]f_0(|p|)d|p|
 \end{align}
So in the repulsive case, we must require
\begin{equation}
\frac{|k|^2}{L^2}  > 4\int_0^{\infty}|p|\left[2\arctanh(v(|p|))-v(|p|)\right]f_0(|p|)d|p|
\end{equation}
 for the repulsive linearized rVP system (with equilibrium $f_0$).

 Thus, requiring \eqref{supcrit_p} in the plasma physics case (i.e. $\sigma = +1$) or \eqref{supcrit_g} in the astrophysical case (i.e. $\sigma = -1$) ensures that $\mathcal{L}[\beta](k, x + 2\pi i y)$ never equals to $1$ on the right half-plane $x \ge 0$ (again, given a spherical, decreasing background).  Hence, we are justified in taking transforms in $t$ of  $\widehat{\rho_t}$ in these cases.

 We mention (as an aside) that the assumption that $f_0$ be \emph{strictly} decreasing in $|p|$ is not essential to the proof going forward.  If $f_0$ is decreasing in $|p|$ and is compactly supported, then our results  continue to remain true.  The only thing that will change is that the support of $\widehat{\beta}^t(k,y)$ in the variable $y$ will be strictly smaller than $|y| \le \frac{|k|}{L}$.  Explicitly, if $f_0$ is supported in $|p| \le P$ for some $P>0$, then the support of $\widehat{\beta}^t$ will be in the set $$|y| \le \frac{|k|}{L}\left( \frac{P}{\sqrt{1+P^2}}\right).  $$  Thus, we would need to evaluate certain quantities at these boundary points rather than at $y=\pm \frac{|k|}{L}$.  Having mentioned this possible extension to decreasing kinetic equilibrium data with compact support, we will stick to the case of strictly decreasing data in the following.

\subsubsection{The Integral Kernel $\mathcal{I}$}

Having established rates of decay in $t$ for the functions $\alpha$ and $\beta$ and determining that we are justified in taking $t$ transforms for sufficiently large $|k|$, we now turn to a closer examination of the integral kernel $\mathcal{I}$.  We begin by noting the following fact:
\begin{eqnarray}
\mathcal{L}[\mathcal{I}(k,\cdot)](2\pi i y) &=& \int_0^{\infty} \mathcal{I}(k,t) e^{-2\pi i y t}dt\nonumber\\
&=& \int_{-\infty}^{\infty} \mathcal{I}(k,t)H(t) e^{-2\pi i y t}dt\nonumber\\
&=& \widehat{\mathcal{I}H}^t(y).
\end{eqnarray}
Thus, for all $t>0$ we can write
\begin{eqnarray}
\mathcal{I}(k,t) &=& \mathcal{I}(k,t)H(t)\nonumber\\
&=& \int_{-\infty}^{\infty}\widehat{\mathcal{I}H}^t(y)e^{2\pi i y t}dy\nonumber\\
&=& \int_{-\infty}^{\infty} \mathcal{L}[\mathcal{I}](k, 2\pi i y)e^{2\pi i y t}dy.
\end{eqnarray}
so long as $\mathcal{L}[\mathcal{I}](k, 2\pi i y)\in L^2(\mathbb{R})$ for a given $k$ (and with the understanding that the integral above represents the extension of the Fourier Transform to $L^2$).  Since we have
\begin{equation*}
\mathcal{L}[\mathcal{I}](k, 2\pi i y)=\frac{\mathcal{L}[\beta](k, 2\pi i y)}{1-\mathcal{L}[\beta](k, 2\pi i y)}
\end{equation*}
from the previous section, our problem is reduced to determining rates of decay for $\mathcal{L}[\beta(k,\cdot)](2\pi i y)$.  Note that our deliberations above show us that the denominator is bounded away from zero by our assumptions on the wavevector $k$.

As above, we have for $t>0$
\begin{align}
\mathcal{L}[\beta](k, 2\pi i y) &= \int_{-\infty}^{\infty} \beta(k,t)H(t) e^{-2\pi i y t}dt\nonumber\\
&=\widehat{\beta}^t \ast \widehat{H}^t (y)\nonumber\\
&=\frac{1}{2}\widehat{\beta}^t(k,y) - \frac{i}{2\pi}\int_{0}^{\infty}\frac{\widehat{\beta}^t(k,y-\tau)-\widehat{\beta}^t(k,y+\tau)}{\tau}d\tau.
\end{align}
We have shown above that the imaginary part of this is compactly supported and the real part is symmetric in $y$ for $|y| \ge \frac{|k|}{L}$.  Hence, for $y \ge \frac{|k|}{L}$ we have
 \begin{eqnarray}
 \mathcal{L}[\beta](k, 2\pi i y) &=& \frac{4\sigma L^2}{|k|^2}\int_0^{\infty}\left[ \frac{yL}{|k|}\arctanh\left(\frac{|k|}{yL}v(|p|)\right)\!-\!v(|p|)\right]\nonumber\\
 &&\;\;\;\;\;\;\;\;\;\;\;\;\;\;\;\;\;\;\;\;\;\;\;\;\;\;\;\;\;\;\;\;\;\;\;\;\;\; \cdot (1+|p|^2)  (-f_0'(|p|))d|p|.
 \end{eqnarray}
 Deliberations in the appendix show that this behaves like $|y|^{-2}$ for large $y$ and so is certainly an $L^2$-function of $y$.

 Given the form of $\mathcal{L}[\mathcal{I}](k, 2\pi i y)$, the fact that $\mathcal{L}[\beta](k, 2\pi i y)$ is in  $L^2$ as a function of $y$, and that $\mathcal{L}[\beta](k, 2\pi i y)$  is strictly bounded away from $1$ by our assumptions, we can conclude that $\mathcal{L}[\mathcal{I}](k, 2\pi i y)$ is also in $L^2$ as a function of $y$ (it is also in $L^1$ as a matter of fact).  Hence, the representation of $\mathcal{I}(k,t)$ given above is valid.

\subsubsection{Decay Rates for Supercritical Modes}

Since we have
\begin{equation*}
\widehat{\rho^h_t}(k) = \alpha(k,t) + \int_0^t \mathcal{I}(k,t-\tau)\alpha(k,\tau)d\tau,
\end{equation*}
and since Assumption \ref{asp 1} implies that $\alpha(k,t)$ will decay $c_k e^{-\epsilon_k |t|^{1/s_k}}$ for some constants $c_k, \epsilon_k>0$ and $ s_k>1$, it remains to examine the decay rate for the convolution integral.  For $t>0$, the Fourier Transform (in $t$) of this quantity is
\begin{align}
\int_{-\infty}^{\infty} H(t)\int_0^t &\mathcal{I}(k,t-\tau)\alpha(k,\tau)e^{-2\pi i \omega t}d\tau dt \nonumber\\
&=\mathcal{L}[\mathcal{I}(k,\cdot)](2\pi i \omega) \cdot \mathcal{L}[\mathcal{\alpha}(k,\cdot)](2\pi i \omega)\nonumber\\
&=\left(\frac{\mathcal{L}[\beta(k,\cdot)](2\pi i \omega)}{1-\mathcal{L}[\beta(k,\cdot)](2\pi i \omega)} \right)\mathcal{L}[\mathcal{\alpha}(k,\cdot)](2\pi i \omega).\label{conv_trans}
\end{align}
Since we are in the supercritical case, the denominator above is bounded away from zero for all $\omega$.

Note that we have
\begin{eqnarray}
\mathcal{L}[\mathcal{\alpha}(k,\cdot)](2\pi i \omega) &=& \widehat{\alpha}^t(k,\cdot)\ast\widehat{H}(\omega) \\
\mathcal{L}[\mathcal{\beta}(k,\cdot)](2\pi i \omega) &=& \widehat{\beta}^t(k,\cdot)\ast\widehat{H}(\omega).
\end{eqnarray}
If we were to take the Fourier transform of these quantities, we would obtain \\$\alpha(k,-t)H(-t)$ and $\beta(k,-t)H(-t)$, respectively.  Assumption \ref{asp 1} ensures that both of these quantities decay like $c_k e^{-\epsilon_k |t|^{1/s_k}}$, and so the Laplace transforms above are elements of $G^{s_k}(\mathbb{R})$ though they will no longer be compactly supported.  In fact, it will not even be the case that these functions are Schwartz class!  For example, our more-or-less explicit form for $\mathcal{L}[\mathcal{\beta}(k,\cdot)](2\pi i \omega)$ in the previous section shows that this quantity only decays like $|\omega|^{-2}$.  Performing analogous computations with $\alpha(k,t)$ shows that this quantity is
\begin{eqnarray}
\widehat{\alpha}^t(k,y) &=&  \frac{2 \pi L }{|k|} \int_{\mathcal{P}(|y|L/|k|)}^{\infty}|p|\sqrt{1+|p|^2} \; \widehat{h}_0(|p|,k)d|p|,
\end{eqnarray}
for $|y| < \frac{|k|}{L}$ and $0$ otherwise.  So for large $|\omega|$,
 \begin{eqnarray}
 \mathcal{L}[\mathcal{\alpha}(k,\cdot)](2\pi i \omega) &=& -\frac{i}{2\pi}\int_0^{\infty}\frac{\widehat{\alpha}^t(k,\omega-\tau)-\widehat{\alpha}^t(k,\omega+\tau)}{\tau}d\tau.
 \end{eqnarray}
 By the apparent symmetry in $y$ for $\widehat{\alpha}^t$, we need only consider the behavior for $\omega \ge |k|/L$.  A calculation similar to the one for $\beta$ gives
 \begin{align}
 \mathcal{L}[\mathcal{\alpha}(k,\cdot)]&(2\pi i \omega) \nonumber\\
 & =\frac{-2iL}{|k|}\int_0^{\infty}\!\!\arctanh\left(\frac{|k|}{L|\omega|}v(|p|)\right)|p|\sqrt{1+|p|^2}\;\widehat{h}_0(|p|,k)d|p|.\label{alpha_trans}
 \end{align}
Hence,  this quantity decays as $|\omega|^{-1}$ for large $|\omega|$.  What we do have is that these functions belong to $G^{s_k}(\mathbb{R})\cap L^2(\mathbb{R})$ by basic properties of the transform.  Moreover, the decay rates quoted above guarantee that
\begin{equation}
\mathcal{L}[\beta(k,\cdot)](2\pi i \cdot) \cdot \mathcal{L}[\alpha(k,\cdot)](2\pi i \cdot) \in G^{s_k}(\mathbb{R})\cap L^1(\mathbb{R})\cap L^2(\mathbb{R}).
\end{equation}
Since we are dealing with a supercritical mode, $1-\mathcal{L}[\beta(k,\cdot)](2\pi i \omega)$ is bounded strictly away from $0$.  Hence, we will certainly have the transform of our convolution integral in $L^2(\mathbb{R})$.  This also shows that $\left(1-\mathcal{L}[\beta(k,\cdot)](2\pi i \omega)\right)^{-1}$ will be in $G^{s_k}(\mathbb{R})$.  As the Laplace transform is bounded strictly away from $1$, the image of any compact set in $\mathbb{R}$ under $\mathcal{L}[\beta(k,\cdot)](2\pi i \cdot)$ yields a compact set of function values where $(1-x)^{-1}$ is analytic.  As $s_k >1$,  $(1-x)^{-1}$ will be in $G^{s_k}$ for these sets.  Since the composition of two Gevrey functions is Gevrey with the maximum degree (see the remark following Proposition 1.4.6 of \cite{R93}) , we see that the denominator above is of Gevrey class $s_k$ as well.  Hence, the Fourier transform of our convolution integral is in $G^{s_k}(\mathbb{R})\cap L^2(\mathbb{R})$ and decays like $|\omega|^{-3}$ for large $|\omega|$.

Now, the calculations given in the appendix below  show that the transform of the convolution integral \eqref{conv_trans} satisfies the assumption of Corollary \ref{Gevrey_cor}.  Hence, the inverse transform of \eqref{conv_trans} will satisfy a sub-exponential decay in time.  Therefore, under Assumption \ref{asp 1}, there exist positive constants $c_k$, $\epsilon_k$ and $s_k>1$ for the supercritical wavevector $k$ so that for $t>0$
\begin{equation}
|\widehat{\rho^h_t}(k)| \le  c_k e^{-\epsilon_k t^{1/s_k}}. \qquad \square
\end{equation}

\subsubsection{Heuristics for Subcritical Modes}

The behavior of the subcritical modes (should there be any) is rather difficult to determine.  In this case, the Fourier-Laplace Transform will definitely have two singularities on the imaginary axis in the plasma-physics case.  In the astrophysics case, there will be two singularities on the real axis by analytic continuation (symmetric with respect to $0$).    A natural idea would be to take the usual Inverse Laplace transform and attempt to evaluate it via an analysis of residues.  If we naively do this, the singularities on the imaginary axis would seem to indicate that subcritical modes are not damped in general for the plasma-physics case (undergoing some kind of oscillatory behavior).  In the astrophysical case, the singularity on the positive real axis would seem to indicate these modes will tend to grow (rapidly) in $t$!

The primary issue with these formal ideas is that the Fourier-Laplace transform of $\beta$ appearing in the transform of the integral kernel fails to be analytic at all points on the imaginary axis whose magnitude is greater than $|k|/L$ (this is intimately related to the fact that the usual branch cut for the $\arctanh$ in the complex plane consists of the two rays on the imaginary axis emanating from $\pm i$).  The singularities for the integral kernel in the plasma physics case are embedded somewhere in this range (or at its endpoints in the critical cases), but analyticity fails at the other points on the imaginary axis  because the imaginary part of the transform for $\beta$ vanishes in this range (while its real part is non-zero).  Note that this is the case for both the repulsive and the attractive interactions.  Thus, it is unclear (at least to the author) whether the inversion formula for the Laplace transform can be used directly in either the attractive or the repulsive regimes for linearized rVP.

\section{Linearized rVP in $\mathbb{R}^3$}\label{full_space_sect}

Note that just as on the torus, we get an independent Volterra equation for each wavevector $\xi$.  By the conservation of the total integral of $h_t$ under the dynamics, we know \emph{a priori} that
\begin{equation}
\widehat{\rho^h_t}(0) = \widehat{\rho^h_0}(0),\label{zero_mode}
\end{equation}
for all times.

As before, the evolution is best analyzed via the Laplace Transform.  Formally taking the Laplace Transform of both sides of \eqref{Volterra2} and solving for the transform of $\widehat{\rho^h_t}$ gives
\begin{equation}
\mathcal{L}[\widehat{\rho^h_{\cdot}}(\xi)](s) = \mathcal{L}[\alpha(\xi,\cdot)](s) + \frac{\mathcal{L}[\beta(\xi,\cdot)](s)}{1-\mathcal{L}[\beta(\xi,\cdot)](s)}\mathcal{L}[\alpha(\xi,\cdot)](s).
\end{equation}
Hence, we have
\begin{equation}
\widehat{\rho^h_t}(\xi) = \alpha(\xi,t) + \int_0^t \mathcal{I}(\xi,t-\tau)\alpha(\xi,\tau)d\tau, \label{soln2}
\end{equation}
with the integral kernel $\mathcal{I}$ defined by
\begin{equation}
\mathcal{L}[\mathcal{I}(\xi,\cdot)](s)=\frac{\mathcal{L}[\beta(\xi,\cdot)](s)}{1-\mathcal{L}[\beta(\xi,\cdot)](s)} \label{kernel2}
\end{equation}
whenever the inversion of the Laplace Transform is justified.

At this point, we can see that our analysis on all of $\mathbb{R}^3$ will closely mirror that on the torus.  Since the wavevector was a fixed parameter in all of our considerations, for any given proof in Section \ref{torus_sect} we need only make the replacement $k/L \to \xi$ to obtain a proof for $\mathbb{R}^3$.  As such, we will not repeat the details.

\subsection*{Proof of Theorem \ref{no_exp_decay_R3}}
As on the torus, Assumption \ref{asp 2} ensures that any decay of $g$ as $t\to \infty$ is the same as the decay for $t\to -\infty$.  Thus the proof of Theorem \ref{no_exp_decay_R3} is precisely the same as it was on the torus.  The same note applies here as well.  The only way to have exponential decay of a given Fourier mode for $t>0$ is that any damping for $t<0$ is strictly slower than exponential (and in particular, the initial data needs to have some asymmetry).

\subsection*{Proof of Theorem \ref{decay_rate_R3}}

The proof of Theorem \ref{decay_rate_R3} follows by exactly the same considerations as on the torus.  Now however, there will definitely be subcritical modes regardless of temperature (simply by the fact that modes are no longer discrete).  Recall that subcritical modes are those associated to non-zero wavevectors not satisfying inequality \eqref{supcrit2_p} or \eqref{supcrit2_g} in the plasma physics or astrophysics case, respectively.

So, in the unbounded space there will always be subcritical modes.  Just as with the torus, we expect that the singularities in both the astrophysical and plasma-physics cases indicate that the corresponding Fourier modes will not decay rapidly in time.  Moreover, these long-lived modes will create huge problems in a full analysis of the behavior of solutions.  As such, determining the exact behavior of solutions on the full space is likely to be a rather difficult problem.

\section{Appendix: Computations Related to $\alpha(k,t)$ and $\beta(k,t)$} \label{Appendix}

We will use Lemma \ref{Gevrey_lem} to show that the transform of the convolution integral \eqref{conv_trans} satisfies the assumptions of Corollary \ref{Gevrey_cor}.  It suffices to show that there is an $R>0$ so that the derivatives of $\mathcal{L}[\alpha(k,\cdot)](2\pi i \omega)$ and $(1-\mathcal{L}[\beta(k,\cdot)](2\pi i \omega))^{-1}$ in $\omega$ are bounded in absolute value for all $|\omega| > R$ and such that the $L^1$-norm of the derivatives of $\mathcal{L}[\beta(k,\cdot)](2\pi i \omega)$ on  $|\omega| > R$ are bounded as in the lemma.

We begin with an examination of $\mathcal{L}[\alpha(k,\cdot)](2\pi i \omega)$.  From the form of this function given in \eqref{alpha_trans}, it is clear that is even in $\omega$ and so we need only concern ourselves with $\omega > R > |k|/L$.  Consider the function
\begin{equation}
f(\omega) = \arctanh\left(\frac{Kv}{L\omega}\right),
\end{equation}
which appears in the integrand.  We take $K, L,$ and $v$ to be fixed, positive parameters for our current discussion (with $0\le v < 1$).  The derivatives of this function can be shown by induction to have the form
\begin{eqnarray}
f^{(2n+1)}(\omega) &=& -\frac{(2n)!KvL^{2n+1}}{(L^2\omega^2-K^2v^2)^{2n+1}}\left(\sum_{i=0}^n C_{i,n-i}^{2n+1}(L^2\omega^2)^{i}(K^2v^2)^{n-i}\right),\\
f^{(2n+2)}(\omega)&=&\frac{(2n+2)!KvL^{2n+3}}{(L^2\omega^2-K^2v^2)^{2n+2}}\omega\left(\sum_{i=0}^{n} C_{i,n-i}^{2n+2}(L^2\omega^2)^{i}(K^2v^2)^{n-i}\right),
\end{eqnarray}
where the non-negative coefficients $C^{m}_{i,j}$ are given by a recurrence relation:
\begin{align*}
C^1_{0,0}&= 1,\;\; C^2_{0,0}= 1,\\
C^{2n+1}_{i,n-i}&= (4n-2i+1)C^{2n}_{i-1,n-i} + (2i+1)C^{2n}_{i,n-1-i},\\
C^{2n+2}_{i,n-i} &= \frac{2n-i+1}{(n+1)(2n+1)}C^{2n+1}_{i,n-i} + \frac{i+1}{(n+1)(2n+1)}C^{2n+1}_{i+1,n-i-1},
\end{align*}
where we interpret any coefficients with negative indices as zero.  The first thing we can conclude is that the absolute value of $f$ and all of its derivatives are strictly decreasing for $\omega \ge R > Kv/L$, and so the $L^{\infty}$-norm in all cases is given by evaluation at $\omega = R$.  We take $R = \sqrt{2}K/L$ for convenience (recall the bound on $v$).  Plugging in this value of $R$ and making obvious estimates gives
\begin{eqnarray}
\left\| f^{(2n)}\right\|_{L^{\infty}([-R,R]^c)} &\le& \left(\frac{\sqrt{2}L}{K}\right)^{2n}(2n)!\sum_{i=0}^{n-1} C_{i,n-1-i}^{2n},\\
\left\| f^{(2n+1)}\right\|_{L^{\infty}([-R,R]^c)} &\le& \left(\frac{\sqrt{2}L}{K}\right)^{2n+1}(2n)!\sum_{i=0}^{n} C_{i,n-i}^{2n+1}.
\end{eqnarray}
Note that since $\arctanh(1/\sqrt{2}) < 1,$ the estimate above holds also for $f$ itself as long as we interpret the ill-defined summation which appears for $n=0$ as $1$.

To estimate the sum of our coefficients, we first note by iterating and making simple (and rather gross) estimates
\begin{eqnarray}
C^{2n}_{i,n-1-i} &\le& 10C^{2n-2}_{i-1,n-2-i}+2C^{2n-2}_{i,n-3-i}+6C^{2n-2}_{i+1,n-4-i},\\
C^{2n+1}_{i,n-i} &\le& 10C^{2n-1}_{i-1,n-i}+2C^{2n-1}_{i,n-1-i}+6C^{2n-1}_{i+1,n-2-i},
\end{eqnarray}
Note that these estimates are reminiscent of the equivalent identities for binomial coefficients.  Using these (rather generous) upper bounds, we see that
\begin{eqnarray}
\sum_{i=0}^{n-1} C_{i,n-1-i}^{2n} &\le& 18 \sum_{i=0}^{n-2} C_{i,n-1-i}^{2n-2} \le (\sqrt{18})^{2n}, \\
\sum_{i=0}^{n} C_{i,n-i}^{2n+1} &\le& 18\sum_{i=0}^{n-1} C_{i,n-i-1}^{2n-1} \le (\sqrt{18})^{2n+1},
\end{eqnarray}
where the final upper bound in each case follows since the first sum in either case is equal to $1$.  Hence, we see that
\begin{eqnarray}
\left\| f^{(2n)}\right\|_{L^{\infty}([-R,R]^c)} &\le& \left(\frac{6L}{K}\right)^{2n}(2n)!,\\
\left\| f^{(2n+1)}\right\|_{L^{\infty}([-R,R]^c)} &\le& \left(\frac{6L}{K}\right)^{2n+1}(2n+1)!.
\end{eqnarray}

Putting all this together, we have for $|\omega| > R =  \sqrt{2}K/L$
\begin{align}
&\left|\frac{d^m}{d\omega^m}\mathcal{L}[\alpha(k,\cdot)](2\pi i \omega)\right|\nonumber\\  &\le\frac{1}{\pi}\int_0^{\infty}\!\!\left|\frac{d^m}{d\omega^m}\arctanh\left(\frac{|k|}{L|\omega|}v(|p|)\right)\right||p|\sqrt{1+|p|^2}\left|\widehat{h_0}(|p|,k)\right|d|p|\nonumber\\
&\le \left(\frac{6L}{K}\right)^{m}(m)!\frac{1}{\pi}\int_0^{\infty}|p|\sqrt{1+|p|^2}\left|\widehat{h_0}(|p|,k)\right|d|p|\nonumber\\
&\le \mathcal{C}\left(\frac{6L}{K}\right)^{m}(m)!,
\end{align}
where the final constant $\mathcal{C}$ is simply given by the remaining integral (which is finite under our assumptions on $h_0$).  Hence,
\begin{equation}
\left\|\frac{d^m}{d\omega^m}\mathcal{L}[\alpha(k,\cdot)](2\pi i \cdot)\right\|_{L^{\infty}([-R,R]^c)} \le \mathcal{C}\left(\frac{6L}{K}\right)^{m}(m)!,
\end{equation}
which is equivalent to the estimate we need to apply Lemma \ref{Gevrey_lem}.  Note that all the derivatives of $\mathcal{L}[\alpha(k,\cdot)](2\pi i \omega)$ in the variable $\omega$ are actually in $L^1([-R,R]^c)$.  However, $\mathcal{L}[\alpha(k,\cdot)](2\pi i \omega)$ itself only decays like $|\omega|^{-1}$ which prevents us from making an $L^1$-estimate for it.  Fortunately, $\mathcal{L}[\beta(k,\cdot)](2\pi i \omega)$ does not suffer from this defect!

For constants $K,L>0$ and any $v \in [0,1) $, consider the function
\begin{equation}
g(\omega) = \frac{L\omega}{K}\arctanh\left(\frac{Kv}{L\omega}\right)-v = \frac{L\omega}{K}f(\omega)-v,
\end{equation}
which appears in the integrand of $\mathcal{L}[\beta(k,\cdot)](2\pi i \omega)$ (c.f. \eqref{Lbetaform}).  Once again, we use the symmetry of the quantities under investigation so that we need only concern ourselves with the region $\omega > K/L$  We have already noted above that for this region this function is strictly decreasing and for large $\omega$ looks like $$ g(\omega) \asymp \frac{K^2v^3}{3L^2\omega^2}.  $$  We play the same game as for the function $f$ above.  The first derivative of this quantity is given by
\begin{equation}
g'(\omega) = \frac{L}{K}\arctanh\left(\frac{Kv}{L\omega}\right)-\frac{vL^2\omega}{L^2\omega^2-K^2v^2},
\end{equation}
which decays (as expected) like
$$ g'(\omega) \asymp -\frac{2K^2v^3}{3L^2\omega^3}.  $$
After the first, the derivatives of $g$ become more predictable:
\begin{eqnarray}
g^{(2n)}(\omega) \!\! &=& \!\! \frac{2(2n-2)!K^2v^3L^{2n}}{(L^2\omega^2-K^2v^2)^{2n}}\left(\sum_{i=0}^{n-1}D^{2n}_{i,n-i-1}(L^2\omega^2)^{i}(K^2v^2)^{n-i-1}\right),\\
g^{(2n+1)}(\omega) \!\! &=& \!\! \frac{-2(2n)!K^2v^3L^{2n+2}\omega}{(L^2\omega^2-K^2v^2)^{2n+1}}\left(\sum_{i=0}^{n-1}D^{2n+1}_{i,n-i-1}(L^2\omega^2)^{i}(K^2v^2)^{n-i-1}\right).
\end{eqnarray}
The recurrence relation for the coefficients is now given by
\begin{align*}
D_{0,0}^2 &= 1\\
D_{i,n-i-1}^{2n+1} &= \frac{(4n-2i)D_{i,n-i-1}^{2n}+(2i+2)D_{i+1,n-i-2}^{2n}}{(2n)(2n-1)},\\
D_{i,n-i}^{2n+2} &=(2i+1)D_{i,n-i-1}^{2n+1}+(4n-2i+3)D_{i-1,n-i}^{2n+1},
\end{align*}
where we once again identify any coefficients with negative subscripts as $0$.

As before, the form of these derivatives makes it clear that they are all decreasing in magnitude for $\omega \ge R =  \sqrt{2}K/L$ (where we use the same $R$ as for the previous function).  Moreover, it is clear by looking at the powers of $\omega$ that $g$ and all of its derivatives are in $L^1([-R,R]^c)$.  Once we estimate the $L^1$-norm of these functions, we will be able to conclude that $\mathcal{L}[\beta(k,\cdot)](2\pi i \omega)$ is in $L^1([-R,R]^c)$ as a function of $\omega$.

For $g$ and its first two derivatives, we can compute explicitly that
\begin{eqnarray}
\left\|g\right\|_{L^1([-R,R]^c)} &=& \frac{K}{2L}\left(2\sqrt{2}v - (2-v^2)\ln\left(\frac{\sqrt{2}+v}{\sqrt{2}-v}\right)\right)\nonumber\\
&\le&\frac{K}{L},\\
\left\|g'\right\|_{L^1([-R,R]^c)} &=&2\sqrt{2}\arctanh\left(\frac{\sqrt{2}v}{2}\right)-2v\nonumber\\
&\le& \frac{1}{2},\\
\left\|g''\right\|_{L^1([-R,R]^c)} &=& \frac{2L}{K}\frac{\sqrt{2}v-(2-v^2)\arctanh(v/\sqrt{2})}{2-v^2}\nonumber\\
&\le& \frac{4L}{K}
\end{eqnarray}
where we have used that $0\le v < 1$ to obtain the upper bounds listed.  After this point, we can use the monotonicity and symmetry of the derivatives along with the formulae given above to give exact values for the $L^1$-norms on $[-R,R]^c$:
\begin{eqnarray}
\left\|g^{2n+1}\right\|_{L^1([-R,R]^c)} &=& 2g^{(2n)}(R)\nonumber\\
&\le& 2(2n-2)!\left(\frac{\sqrt{2}L}{K}\right)^{2n}\sum_{i=0}^{n-1}D^{2n}_{i,n-i-1}\nonumber\\
&\le& 2\left(\frac{6\sqrt{2}L}{K}\right)^{2n}(2n-2)!,\\
\left\|g^{2n+2}\right\|_{L^1([-R,R]^c)} &=& -2g^{(2n+1)}(R)\nonumber\\
&\le& 2(2n)!\left(\frac{\sqrt{2}L}{K}\right)^{2n+1}\sum_{i=0}^{n-1}D^{2n+1}_{i,n-i-1}\nonumber\\
&\le& 2\left(\frac{2\sqrt{13}L}{K}\right)^{2n+1}(2n)!.
\end{eqnarray}
Once again, we have used rather gross estimates to obtain
\begin{eqnarray}
D^{2n+1}_{i,n-i-1} &\le& 6D^{2n-1}_{i-1,n-i-1}+10D^{2n-1}_{i,n-i-2}+10D^{2n-1}_{i+1,n-i-3},\\
D^{2n+2}_{i,n-i} &\le& 21D^{2n}_{i-1,n-i}+8D^{2n}_{i,n-i-1}+6D^{2n}_{i+1,n-i-2},
\end{eqnarray}
and so the summation of these coefficients is bounded by the appropriate power of $2\sqrt{21}$ (again, since the first summation yields $1$).

Using these estimates, we see that \eqref{Lbetaform} now yields
\begin{align}
&\left\|\frac{d^m}{d\omega^m}\mathcal{L}[\beta(k,\cdot)](2\pi i \cdot)\right\|_{L^1([-R,R]^c)} \nonumber\\
& \;\;\;\;\;\; \le \frac{4 L^2}{|k|^2}\int_0^{\infty}\left\|g^{(m)}\right\|_{L^1([-R,R]^c)} (1+|p|^2)  (-f_0'(|p|))d|p|\nonumber\\
& \;\;\;\;\;\; \le \mathcal{C} \left(\frac{6\sqrt{2}L}{|k|}\right)^{m+1}(m+1)!,
\end{align}
which is precisely the sort of estimate needed in Lemma \ref{Gevrey_lem} (with $m_0=1$).

Finally, we must concern ourselves with the remaining factor appearing in the transform of the integral kernel: $$ \frac{1}{1-\mathcal{L}[\beta(k,\cdot)](2\pi i \omega)}. $$  By Lemma \ref{Gevrey_lem} and our previous results, we need only show that this function and all of its derivatives have appropriate bounds in $L^{\infty}([-R,R]^c)$.  Note that we only need these estimates in cases when the denominator of this expression is bounded strictly away from zero.  Let $\epsilon$ either be the minimum value of the denominator in the expression above for $\omega \in [-R,R]^c$ or equal to $1$ if this minimum is greater than $1$ (so that $0 < \epsilon \le 1$).

For any set of indices such that $j_1+j_2 \cdots +j_{i+1}=n$, the multinomial coefficient is given by
$$\left(\begin{array}{c}n\\j_1,j_2,\ldots ,j_{i+1}\end{array}\right) = \frac{n!}{j_1!j_2!\cdots j_{i+1}!}.  $$
Using these coefficients, the derivatives of the generic function $h(x) = (1-g(x))^{-1}$ can be written succinctly as
\begin{align*}
h^{(n)}(x) =  \sum_{i=0}^{n-1} & (1-g(x))^{-(i+2)}\\
&\cdot \left[\sum_{j_1+j_2 \cdots +j_{i+1}=n}\left(\begin{array}{c}n\\j_1,j_2,\ldots ,j_{i+1}\end{array}\right)g^{(j_1)}(x)g^{(j_2)}(x) \cdots g^{(j_{i+1})}(x)\right],
\end{align*}
via the Fa\`a di Bruno formula (with the understanding that all of the indices \\ $j_1, j_2, \ldots j_{i+1}$ are non-zero).

So, we see
\begin{align}
&\left|\frac{d^m}{d\omega^m}\frac{1}{1-\mathcal{L}[\beta(k,\cdot)](2\pi i \omega)}\right|\epsilon^{m+1}\\
&\le\sum_{i=0}^{m-1}\cdot \left[\sum_{j_1+j_2 \cdots +j_{i+1}=m}\left(\begin{array}{c}m\\j_1,j_2,\ldots ,j_{i+1}\end{array}\right)|g^{(j_1)}(\omega)||g^{(j_2)}(\omega)| \cdots |g^{(j_{i+1})}(\omega)|\right],\nonumber
\end{align}
where $g(\omega) = \mathcal{L}[\beta(k,\cdot)](2\pi i \omega)$.  Using the monotonicity properties we developed above, we know that
\begin{equation}
\left\|g^{(j_k)}\right\|_{L^{\infty}([-R,R]^c)} \le \mathcal{C}\left(\frac{6\sqrt{2}L}{|k|}\right)^{j_k}j_k!,
\end{equation}
where we can assume that $\mathcal{C}\ge 1$.  Thus,
\begin{align}
&\left|\frac{d^m}{d\omega^m}\frac{1}{1-\mathcal{L}[\beta(k,\cdot)](2\pi i \omega)}\right|\nonumber\\
& \;\;\;\; \le \mathcal{C}\epsilon^{-1}\left(\frac{6\sqrt{2}L\mathcal{C}}{\epsilon |k|}\right)^{m}m!\cdot\left(\textrm{number of partitions of } m\right).
\end{align}
Using the well-known result (c.f. \cite{E42})
\begin{equation}
p(m) = \textrm{number of partitions of } m \asymp \frac{1}{4\sqrt{3}m}\exp\left(\pi \sqrt{\frac{2m}{3}}\right),
\end{equation}
we see that there is a constant $\widetilde{\mathcal{C}}$ so that
\begin{equation}
\left\|\frac{d^m}{d\omega^m}\frac{1}{1-\mathcal{L}[\beta(k,\cdot)](2\pi i \cdot)}\right\|_{L^{\infty}([-R,R]^c)} \le \widetilde{\mathcal{C}}\left(\frac{6\sqrt{2}L\mathcal{C}}{\epsilon |k|}\exp\left(\pi \sqrt{\frac{2}{3}}\right)\right)^{m}m!.
\end{equation}

Multiplying these three functions together gives the transform of our convolution integral, \eqref{conv_trans}.  Hence, Lemma \ref{Gevrey_lem} implies that this transform satisfies the requirements of Corollary \ref{Gevrey_cor}  (with $m_0=1$). As such, the convolution integral itself has the sub-exponential decay guaranteed by the corollary.

\end{document}